\documentclass[10pt,osajnl2,twocolumn,showpacs]{revtex4}
\pdfoutput=1

\usepackage[utf8]{inputenc}

\usepackage{graphicx}
\graphicspath{{./}{./figs/}}

\usepackage{amsmath,amsfonts,amssymb}
\usepackage{xspace,lastpage}
\usepackage[dvipsnames,hyperref]{xcolor}
\colorlet{myLinkColor}{MidnightBlue}

\newcommand{\TheTitle}{Exploiting spatial sparsity for multi-wavelength imaging in
  optical interferometry}
\newcommand{\AffiliationNumber}[1]{}
\newcommand{\TheAuthors}{%
  Éric Thiébaut\AffiliationNumber{1},
  Ferréol Soulez\AffiliationNumber{1} and
  Loïc Denis\AffiliationNumber{2}}

\usepackage{hyperref}
\hypersetup{
  pdftitle = {\TheTitle},
  pdfauthor = {\TheAuthors},
  breaklinks=true,
  colorlinks=true,
  linkcolor=myLinkColor,
  citecolor=myLinkColor,
  filecolor=myLinkColor,
  menucolor=myLinkColor,
  urlcolor=myLinkColor,
}

\newcommand{\myURL}[1]{\href{#1}{\texttt{#1}}}

\makeatletter
\newcommand{\MathFuncName}[1]{{\operator@font #1}}
\newcommand{\MathFunc}[1]{\mathop{\operator@font #1}\nolimits}
\newcommand{\MathFuncWithLimits}[1]{\mathop{\operator@font #1}\limits}
\makeatother

\newcommand{\Tag}[1]{\mathsf{#1}}        
\newcommand{\V}[1]{\boldsymbol{#1}}      
\newcommand{\M}[1]{\mathbf{#1}}          
\newcommand{\TransposeLetter}{\top}
\newcommand{\T}{^{\TransposeLetter}}     
\newcommand{\sgn}{\MathFunc{sgn}}        
\newcommand{\Cov}{\MathFunc{Cov}}        
\newcommand{\Card}{\MathFunc{Card}}      
\newcommand{\prox}{\operatorname{prox}}  
\newcommand{\proxy}[1]{\widetilde{#1}}   

\newcommand{\Eq}[1]{Eq.~(\ref{#1})}

\newcommand{\YORICK}[1]{}

\newcommand{\PDer}[2]{\frac{\partial #1}{\partial #2}}

\newcommand{\norm}[1]{\Vert #1\Vert}
\newcommand{\Norm}[1]{\left\Vert #1\right\Vert}
\newcommand{\bigNorm}[1]{\bigl\Vert #1\bigr\Vert}

\newcommand{\abs}[1]{\vert #1\vert}
\newcommand{\Abs}[1]{\left\vert #1\right\vert}

\newcommand{\argmin}{\MathFuncWithLimits{arg\,min}}

 \newcommand{\EndProof}{\ensuremath{\blacksquare}}

\newcommand{\FT}[1]{\hat{#1}}            

\newcommand{\Set}[1]{\mathbb{#1}}
\newcommand{\Reals}{\mathbb{R}}

\newcommand{\bydef}{\stackrel{\text{\tiny def}}{=}}

\newcommand{\eg}{\emph{e.g.}\xspace}
\newcommand{\ie}{\emph{i.e.}\xspace}

\newcommand{\etc}{\emph{etc.}\xspace}

\newcommand{\textsup}[1]{\ensuremath{^{\text{#1}}}}

\newcommand{\DataTag}{\Tag{data}}
\newcommand{\JointTag}{\Tag{joint}}
\newcommand{\PriorTag}{\Tag{prior}}
\newcommand{\SparseTag}{\Tag{sparse}}

\newcommand{\SpectralTag}{\Tag{spectral}}

\newcommand{\Fcost}{f}
\newcommand{\Fdata}{\Fcost_\DataTag}
\newcommand{\Fjoint}{\Fcost_\JointTag}
\newcommand{\Fprior}{\Fcost_\PriorTag}
\newcommand{\Fsparse}{\Fcost_\SparseTag}

\newcommand{\Fspectral}{\Fcost_\SpectralTag}

\newcommand{\BSMEM}{\textsc{Bsmem}\xspace}
\newcommand{\Clean}{\textsc{Clean}\xspace}
\newcommand{\Mira}{Mi\textsc{ra}\xspace}

\newcommand{\Wisard}{\textsc{Wisard}\xspace}
\newcommand{\Yorick}{\textsc{Yorick}\xspace}
\newcommand{\Lasso}{\textsc{Lasso}\xspace}
\newcommand{\Gravity}{\textsc{Gravity}\xspace}

\newcommand{\YorickCode}[1]{}

\definecolor{FireBrick}{rgb}{0.70,0.13,0.13}

\begin{document}

\title{\TheTitle}

\renewcommand{\AffiliationNumber}[1]{\ensuremath{^{#1}}}

\author{\TheAuthors}

\affiliation{\AffiliationNumber{1} Université de Lyon, Lyon, F-69003, France;
  Université Lyon 1, Observatoire de Lyon, 9 avenue Charles André, Saint-Genis
  Laval, F-69230, France; CNRS, UMR 5574, Centre de Recherche Astrophysique de
  Lyon; École Normale Supérieure de Lyon, Lyon, F-69007,
  France.\\
  \AffiliationNumber{2} Université de Lyon, F-42023, Saint-Etienne, France,\\
  CNRS, UMR5516, Laboratoire Hubert Curien, F-42000, Saint-Etienne, France,\\
  Université de Saint-Etienne, Jean Monnet, F-42000, Saint-Etienne, France.}


\begin{abstract}
Optical interferometers provide multiple wavelength measurements.  In order to
fully exploit the spectral and spatial resolution of these instruments, new
algorithms for image reconstruction have to be developed.  Early attempts to
deal with multi-chromatic interferometric data have consisted in recovering a
gray image of the object or independent monochromatic images in some spectral
bandwidths.  The main challenge is now to recover the full 3-D
(spatio-spectral) brightness distribution of the astronomical target given all
the available data.  We describe a new approach to implement multi-wavelength
image reconstruction in the case where the observed scene is a collection of
point-like sources.  We show the gain in image quality (both spatially and
spectrally) achieved by globally taking into account all the data instead of
dealing with independent spectral slices.  This is achieved thanks to a
regularization which favors spatial sparsity and spectral grouping of the
sources.  Since the objective function is not differentiable, we had to
develop a specialized optimization algorithm which also accounts for
non-negativity of the brightness distribution.
\end{abstract}


\maketitle


\section{Introduction}
\label{sec:introduction}

The objective of stellar interferometric imaging is to recover an
approximation of the specific brightness distribution $I_\lambda(\V{\theta})$
of observed astronomical objects given measurements providing incomplete
samples of the spatial Fourier transform of $I_\lambda(\V{\theta})$.
Reconstruction of a monochromatic image from optical interferometry data is a
challenging task which has been the subject of fruitful research and resulted
in various algorithms (\eg, \Mira \cite{Thiebaut-2008-Marseille}, \BSMEM
\cite{Buscher-1994-BSMEM,Baron_Young-2008-Marseille}, \Wisard
\cite{Meimon_et_al-2005-weak_phase_imaging}, the \emph{building-block method}
\cite{Hofmann_Weigelt-1993-building_blocks}).  When dealing with
multi-spectral data, a first possibility is to process each wavelength
independently and reconstruct a monochromatic image for each subset of
measurements from a given spectral channel.  For instance, this is what have
been done by le~Bouquin et al. \cite{leBouquin_et_al-2009-TLep} for the
multi-spectral images of the Mira star T~Lep.  Another possibility is to
exploit some assumed spectral continuity of $I_\lambda(\theta)$ and process
the multi-spectral data globally to reconstruct an approximation of the 3-D
distribution $I_\lambda(\theta)$.  This computationally more challenging
approach can potentially lead to better reconstructions. Significant
improvements have been shown when following such spatio-spectral processing in
the context of integral field spectral spectroscopy
\cite{Soulez_et_al-2011-Lisbon,
  Bongard_et_al-2011-hyperspectral_deconvolution, Bourguignon2011}.  This
paper describes a method to jointly reconstruct multi-spectral optical
interferometric data.

In order to simplify the problem, we restricted our study to the cases where
the complex visibilities are observed and where the observed scene is a
collection of point-like sources. This correspond, for instance, to the
science case of the instrument \Gravity which will be installed at the Very
Large Telescope Interferometer (VLTI) to carry out astrometry with absolute
phase reference of stars in the galactic center or in globular clusters
\cite{Gillessen_et_al-2010-GRAVITY}. In some sense, this latter assumption
makes our algorithm a successor of the \Clean algorithm
\cite{Hogbom-1974-CLEAN, Schwarz-1978-Clean} and the \emph{building-block
  method} \cite{Hofmann_Weigelt-1993-building_blocks} developed for recovering
monochromatic images from radio and optical interferometric data respectively.
The \Clean algorithm have been proposed for the processing of \Gravity
interferometric data \cite{Vincent_el_al-2011-hotspot} but only considering
``gray'' data and not more than three stars in the field of view.  In addition
to processing multi-variate data, we also introduce the explicit minimization
of a non-differentiable regularization term so as to favor spatial sparsity of
the reconstructed brightness distribution in a way which is known to be more
efficient \cite{Donoho-2006-ell1_approx, Fornasier_Rauhut-2008-joint_sparsity}
than greedy algorithms like \Clean
\cite{Marsh_Richardson-1987-CLEAN_objective_function} or the
\emph{building-block method} \cite{Hofmann_Weigelt-1993-building_blocks}.  The
method presented in this paper improves over early developments presented as
an invited paper at the 2012 SPIE Conf. on Astronomical Telescopes \&
Instrumentation in Amsterdam\cite{Thiebaut_Soulez-2012-Amsterdam}.

Our paper is organized as follows: we first summarize the inverse approach for
image reconstruction from interferometric data and discuss various
possibilities to impose spatial sparsity, we then detail our algorithm for
minimizing the objective function; finally we present some results on
simulated data and discuss the advantages of our approach.

\section{Method}
\label{sec:method}

\subsection{General principle of image reconstruction}

Following an inverse approach, we state image reconstruction as a constrained
optimization problem \cite{Thiebaut-2009-interferometric_imaging}:
\begin{equation}
  \V{x}^{+} = \argmin_{\V{x} \in \Set{X}}
  \left\{ \Fdata(\V{x}) + \mu \, \Fprior(\V{x}) \right\}
  \label{eq:image-problem}
\end{equation}
where $\V{x} \in \Set{R}^{\abs{\V{x}}}$ are the sought image parameters,
$\abs{\V{x}} = \Card(\V{x})$ is the number of parameters, $\Set{X} \subset
\Set{R}^{\abs{\V{x}}}$ is the subset of feasible parameters, $\Fdata(\V{x})$
is a data fitting term which enforces agreement of the model with the
measurements $\V{y} \in \Set{R}^{\abs{\V{y}}}$, $\Fprior(\V{x})$ is a
regularization term and $\mu > 0$ is a so-called hyper-parameter used to tune
the relative weight of the regularization.  Following a Bayesian
interpretation, $\Fdata$ is the opposite of the log-likelihood of the
measurements, $\Fprior$ is the opposite of the log of the prior distribution
of the parameters, and $\V{x}^{+}$ defines the maximum a posteriori (MAP)
estimate.

Constraining the solution to belong to the feasible set $\Set{X}$ is a mean
to impose \emph{strict constraints} such as the non-negativity:
\begin{equation}
  \Set{X} = \{\V{x} \in \Set{R}^{n} : \V{x} \ge 0\}
  \label{eq:feasible-set}
\end{equation}
where the inequality $\V{x} \ge 0$ is to be taken element-wise.

\subsection{Direct model and likelihood}

In this study, we assume that the optical interferometric data consist in
Fourier transform of the brightness distribution $\FT{I}_{\lambda}(\V{\nu})$
measured for a finite set of spatial frequencies $\V{\nu} = \V{B}/\lambda$
with $\V{B}$ the interferometric baseline (projected in a plane perpendicular
to the line of sight) and $\lambda$ the wavelength
\cite{Thiebaut_Giovannelli-2010-interferometry}.

The most practical representation of a multi-variate distribution such as
$I_\lambda(\theta)$ by a finite number of parameters $\V{x}$ consists in
sampling $I_\lambda(\theta)$ separately along its spatial and spectral
dimensions.  The image parameters are then:
\begin{equation}
  x_{n,\ell} \approx I_{\lambda_\ell}(\V{\theta}_n)
  \label{eq:multi-spectral-parameters}
\end{equation}
for $\lambda_\ell \in \Set{W}$ the list of sampled wavelengths and
$\V{\theta}_n \in \Set{A}$ the list of angular directions, the so-called
\emph{pixels}.

For the sake of notational simplicity, we use the same wavelengths in $\Set{W}$ as the
ones of the data spectral channels and we denote by $y_{p,m,\ell}$ the real
($p=1$) or imaginary ($p=2$) part of the complex visibility obtained with
$m$th baseline in $\ell$th spectral channel.  This notation is intended to
clarify the equations and does not impose or assume that all baselines have
been observed in all spectral channels.  By considering that complex numbers
are just pairs of real values, our notation also avoids dealing with complex
arithmetic.  In our framework, the model of the data is affine:
\begin{align}
  y_{p,m,\ell}
  &= (\M{H}\cdot\V{x})_{p,m,\ell} + e_{p,m,\ell} \notag \\
  &= \sum\nolimits_{n} H_{p,m,n,\ell} \, x_{n,\ell} + e_{p,m,\ell} \, ,
  \label{eq:direct-model}
\end{align}
where the term $\V{e}$ accounts for noise and modeling approximations.
Formally, the coefficients of the operator $\M{H}$ are given by
\cite{Thiebaut_Giovannelli-2010-interferometry}:
\begin{equation}
  H_{p,m,n,\ell} =
  \left\{\begin{array}{ll}
  {\displaystyle + \cos(\V{\theta}_n\T\!\!\cdot\V{B}_{m}/\lambda_\ell)}
  & \text{for $p=1$}\\[1ex]
  {\displaystyle - \sin(\V{\theta}_n\T\!\!\cdot\V{B}_{m}/\lambda_\ell)}
  & \text{for $p=2$}\\
  \end{array}\right.
  \label{eq:multi-spectral-model-matrix}
\end{equation}
with $\V{B}_m$ the $m$th observed baseline and $\V{\theta}_n\T\!\!\cdot\V{B}_{m}$
the usual scalar product between $\V{B}_{m}$ and $\V{\theta}_n$.

At least because of the strict constraints imposed by the feasible set
$\Set{X}$, solving the image reconstruction problem in \Eq{eq:image-problem}
must be carried out by an iterative algorithm.  Owing to the size of the
problem, a fast version of $\M{H}$ has to be implemented.  First, we note that
the model is separable along the spectral dimension (using a conventional
matrix representation, $\M{H}$ would have a block diagonal structure):
\begin{equation}
  \V{y}_{\ell} = \M{H}_{\ell} \cdot \V{x}_{\ell} + \V{e}_{\ell} \, ,
  \label{eq:direct-model-in-a-channel}
\end{equation}
where the index $\ell$ denotes the sub-vector or the sub-operator restricted
to the coefficients corresponding to the $\ell$th spectral channel.  With the
generalization of multi-processor computers or multi-core processors, this
property of the operator $\M{H}$ may be easily exploited to parallelize the
code to apply $\M{H}$ (or its adjoint $\M{H}\T$) to a given argument.  Second,
an algorithm such as the \emph{nonuniform fast Fourier transform} (NU-FFT)
\cite{Fessler_Sutton-2003-NUFFT} can be implemented to speed up the
computations by approximating the operator $\M{H}_{\ell}$ by:
\begin{equation}
  \M{H}_{\ell} \approx \M{R}_{\ell} \cdot \M{F} \cdot \M{S}
  \label{eq:multi-spectral-nufft}
\end{equation}
where $\M{F}$ is the discrete Fourier transform (DFT), $\M{R}_{\ell}$
interpolates the discrete spatial frequencies resulting from the DFT at the
frequencies observed in $\ell$th channel and $\M{S}$ is a zero-padding and
apodizing operator.  Zero-padding improves the accuracy of the approximation,
while apodization pre-compensates for the convolution by the interpolation
kernel used in $\M{R}_{\ell}$ \cite{Fessler_Sutton-2003-NUFFT}.  Note that
only the interpolation in the Fourier domain $\M{R}_{\ell}$ depends on the
spectral channel.  In NU-FFT, $\M{S}$ is diagonal, $\M{R}_{\ell}$ is very
sparse and $\M{F}$ is implemented by a fast Fourier transform (FFT) algorithm,
thus the approximation in \Eq{eq:multi-spectral-nufft} is very fast to
compute.

Assuming Gaussian noise distribution, the likelihood term writes:
\begin{equation}
  \Fdata(\V{x}) = \frac{1}{2} \, (\M{H} \cdot \V{x} - \V{y})\T
  \cdot \M{W} \cdot (\M{H} \cdot \V{x} - \V{y})
  \label{eq:fdata}
\end{equation}
where $\M{W} \in \Set{R}^{\abs{\V{y}} \times \abs{\V{y}}}$ is a statistical
weighting matrix; in principle, $\M{W}$ is the inverse of the covariance
matrix of the measurements: $\M{W} = \Cov(\V{y})^{-1}$.

\subsection{Regularization based on spatial sparsity}

Due to the voids in the spatial frequencies covered by the observations, the
constraints provided by the data alone do not suffice to define a unique
image.  The prior constraints imposed by $\Fprior(\V{x})$ are then required to
help choosing a unique solution among all the images that are compatible with
the measurements.

In this paper, we focus on a particular type of astronomical targets which
consist in a number of point-like sources with different spectral energy
distributions.  This includes the case of multiple stars, globular clusters,
or groups of stars as observed in the center of our galaxy.  For such objects,
the most effective means to regularize the problem is to favor spatially
sparse distributions, \ie images with as few sources as possible on a dark
background.  In this section, we derive expressions of the regularization term
$\Fprior(\V{x})$ suitable to favor spatially sparse distributions.

\subsubsection{Fully separable sparsity}

It is now well known that using the $\ell_1$ norm as the regularization term
is an effective mean to impose the sparsity of the solution while
approximating the data \cite{Donoho-2006-ell1_approx}.  This leads to take
$\Fprior(\V{x}) = \Fsparse(\V{x})$ with:
\begin{equation}
  \Fsparse(\V{x}) = \norm{\V{x}}_1 \bydef \sum_{k,\ell} \Abs{x_{k,\ell}}
  = \sgn(\V{x})\T\!\!\cdot\V{x} \, ,
  \label{eq:sparse-prior}
\end{equation}
where $\sgn(\V{x})$ is the sign function applied element-wise to the
parameters $\V{x}$.  When the parameters are non-negative, $\sgn(\V{x}) =
\V{1}$ with $\V{1} \bydef (1,\ldots,1)\T$.

The regularization term $\Fsparse(\V{x})$ in \Eq{eq:sparse-prior} is
completely separable.  In our framework where the model is spectrally
separable, the global criterion defined in \Eq{eq:image-problem} is therefore
separable along the spectral dimension.  Provided data from different spectral
channels are statistically independent, the image reconstruction can be solved
independently for each spectral channel.

If the wavelength samples $\lambda_\ell$ of the discrete model $\V{x}$ of
$I_\lambda(\theta)$ do not coincide with the effective wavelengths of the
data, spectral interpolation of the model is required to match the observed
wavelengths.  In this case, a certain spectral correlation is intrinsic to the
model and the 3-D image reconstruction has to be performed globally even if
the regularization does not impose any kind of spectral continuity.

\subsubsection{Non-separable spatial-only sparsity}

Physically, sources emit light at all wavelengths and we expect better image
reconstruction if we can favor restored sources having the same position
whatever the wavelength.  Clearly, this is not achieved by the regularization
$\Fsparse(\V{x})$ in \Eq{eq:sparse-prior} which is fully separable.  In
order to impose some spectral continuity while favoring spatial sparsity, we
consider the following regularization instead
\cite{Fornasier_Rauhut-2008-joint_sparsity, Soulez_et_al-2011-Lisbon}:
\begin{equation}
  \Fjoint(\V{x}) = \sum\nolimits_{n}
  \left(\sum\nolimits_{\ell} x_{n,\ell}^2\right)^{\!1/2}
  \label{eq:joint-sparse-prior}
\end{equation}
with $n$ the spatial index (pixel) and $\ell$ the spectral channel.  The fact
that such a regularization favors spatial sparsity and spectral grouping is a
consequence of the triangular inequality
\cite{Fornasier_Rauhut-2008-joint_sparsity}.  This sparsity prior is a special
case of several recent generalizations such as group \Lasso
\cite{yuan2006model}, mixed norms \cite{Kowalski2009mixednorms} or structured
sparsity \cite{jenatton2011structured}.

\subsubsection{Explicit spectral continuity and gray object}

The penalization defined in \Eq{eq:joint-sparse-prior} can be seen as a
spatial regularization term which favors spectral grouping but no spectral
continuity nor spectral smoothness.  Some authors
\cite{Thiebaut_Mugnier-2006-nulling, Soulez_et_al-2011-Lisbon,
  Bongard_et_al-2011-hyperspectral_deconvolution} have shown the efficiency of
exploiting the spectral continuity of the sought distribution by using, in
addition to a spatial regularization term, an additional spectral
regularization like:
\begin{equation}
  \Fspectral(\V{x}) = \sum_n \mu_n
  \sum_\ell (x_{n,\ell+1} - x_{n,\ell})^2
  \label{eq:spectral-regul}
\end{equation}
with $\mu_n > 0$ suitable regularization weights.  In the limit $\mu_n
\rightarrow \infty, \forall n$, the regularization in \Eq{eq:spectral-regul}
amounts to assuming that the spectral energy distributions of all sources are
flat, that is:
\begin{equation}
  x_{n,\ell} = g_{n} \,, \quad \forall (n, \ell)
  \label{eq:gray-image}
\end{equation}
where $\V{g}$ is a \emph{gray image} of the object which does not depend on
the spectral index $\ell$.  To speed up the reconstruction, only the gray
image has to be reconstructed, using the model:
\begin{equation}
  y_{p,m,\ell} = \sum_{n} H_{p,m,n,\ell} \, g_{n} + e_{p,m,\ell} \, ,
  \label{eq:gray-model}
\end{equation}
and, to impose the spatial sparsity, the regularization on $\V{g}$ is:
\begin{equation}
  \label{eq:gray-regularization}
  \Fsparse(\V{g}) = \norm{\V{g}}_1 = \sum\nolimits_n \abs{g_n}
  = \sgn(\V{g})\T\!\!\cdot\V{g} \, .
\end{equation}

\subsection{Optimization algorithm}

Most existing image reconstruction algorithms for optical interferometry (\eg,
\Mira \cite{Thiebaut-2008-Marseille}, \BSMEM \cite{Buscher-1994-BSMEM,
  Baron_Young-2008-Marseille} and \Wisard
\cite{Meimon_et_al-2005-weak_phase_imaging}) were designed for minimizing a
smooth cost function.  For that purpose, non-linear conjugate gradient method
\cite{Nocedal_Wright-2006-numerical_optimization} or limited memory
quasi-Newton methods such as VMLM-B \cite{Thiebaut-2002-optim_bdec} are quite
efficient and easy to use as they only require computing the cost function and
its gradient.  A notable exception is
\textsc{Macim}\cite{Ireland_et_al-2006-MACIM} which is based on a
Markov-Chain-Monte-Carlo (MCMC) optimization strategy suitable, in theory, for
any type of criteria, in particular the non-smooth and non-convex ones; in
practice, this is however too computationally intensive for estimating a large
number of parameters as it is the case for image reconstruction. When using
non smooth regularizations as the ones in \Eq{eq:sparse-prior} and
\Eq{eq:joint-sparse-prior} to impose spatial sparsity, optimization algorithms
based on Newton method (that is, on a quadratic approximation of the cost
function) are inefficient and completely different optimization strategies
must be followed to solve the problem in \Eq{eq:image-problem} with
non-differentiable cost functions.  In our algorithm, we introduce variables
splitting \cite{Combettes_Pesquet-2011-proximal_methods} to handle the two
terms of the cost function as independently as possible and we implement an
\emph{alternating direction method of multipliers}
\cite{Boyd_et_al-2010-method_of_multipliers} (ADMM) to solve the resulting
constrained problem.  The augmented Lagrangian with ADMM emerges as the most
effective in the family of decomposition methods that includes proximal
methods \cite{Combettes_Pesquet-2011-proximal_methods}, variable splitting
with quadratic penalty \cite{wang2008new}, iterative Bregman
\cite{yin2008bregman}.  See \cite{Afonso_et_al-2011-C_SALSA} for detailed
comparisons.

\subsubsection{Variable Splitting and ADMM}
\label{sec:ADMM}

Introducing auxiliary variables $\V{z}$, minimization of the two-term cost
function in \Eq{eq:image-problem} can be recast in the equivalent constrained
problem:
\begin{equation}
  \min_{\V{x} \in \Set{X}, \V{z}} \left\{ \Fdata(\V{z})
  + \mu\,\Fprior(\V{x}) \right\}
  \quad \text{s.t.}\quad \V{x} = \V{z} \, .
  \label{eq:split-problem}
\end{equation}
Imposing that $\V{x} \in \Set{X}$ (\ie $\V{x} \ge \V{0}$) rather than $\V{z}
\in \Set{X}$ is not arbitrary and our motivation for that choice is explained
in what follows.  Another possible splitting would have been to choose the
auxiliary variables as $\V{z} = \M{H}\cdot\V{x}$ but this would have prevented
us to exploit the separability of the resulting penalty with respect to
$\V{x}$.

The augmented Lagrangian \cite{Nocedal_Wright-2006-numerical_optimization} is
a very useful method to deal with constrained problems such as the one in
\Eq{eq:split-problem}.  In our case, the augmented Lagrangian writes:
\begin{align}
  \mathcal{L}_{\rho}(\V{x},\V{z},\V{u}) &= \Fdata(\V{z})
  + \mu\,\Fprior(\V{x}) \notag \\
  &\quad + \V{u}\T\!\!\cdot(\V{x} - \V{z})
  + \frac{\rho}{2} \, \norm{\V{x} - \V{z}}_2^2 \, ,
  \label{e:augmented-Lagrangian}
\end{align}
with $\V{u}$ the Lagrange multipliers associated with the constraint $\V{x} =
\V{z}$, $\rho > 0$ the quadratic weight of the constraints, and
$\norm{\V{v}}_2$ the Euclidean ($\ell_2$) norm of $\V{v}$.  Note that taking
$\rho=0$ yields the classical Lagrangian of the constrained problem.

The alternating direction method of multipliers
\cite{Boyd_et_al-2010-method_of_multipliers} (ADMM) consists in alternatively
minimizing the augmented Lagrangian for $\V{x}$ given $\V{z}$ and $\V{u}$,
then for $\V{z}$ given $\V{x}$ and $\V{u}$, and finally updating the
multipliers $\V{u}$.  This scheme, adapted to our specific problem in
\Eq{eq:split-problem}, is detailed by the following algorithm with the
convention that $\V{v}^{(t)}$ is the value of $\V{v}$ at iteration number $t$:

\smallskip

\textbf{Algorithm 1.}  \emph{Resolution of problem~(\ref{eq:split-problem}) by
  alternating direction method of multipliers}.  Choose initial variables
$\V{z}^{(0)}$ and Lagrange multipliers $\V{u}^{(0)}$.  Then repeat, for $t =
1, 2, \ldots$ until convergence, the following steps:
\begin{align}
  \intertext{1.~choose $\rho^{(t)} > 0$ and update variables $\V{x}$:}
  \V{x}^{(t)}
  &= \argmin_{\V{x} \in \Set{X}}
  \mathcal{L}_{\rho^{(t)}}\bigl(\V{x}, \V{z}^{(t-1)}, \V{u}^{(t-1)}\bigr) \notag \\
  &= \argmin_{\V{x} \in \Set{X}}
  \Bigl\{ \Fprior(\V{x})
  + \frac{\rho^{(t)}}{2\,\mu} \, \bigNorm{ \V{x} - \proxy{\V{x}}^{(t)} }_2^2 \Bigr\}
  \label{eq:x-update}
  \intertext{\hspace*{2.5ex}with:}
  \proxy{\V{x}}^{(t)} &= \V{z}^{(t-1)} - \V{u}^{(t-1)}/\rho^{(t)} \, ;
  \label{eq:x-proxy-def}
  \intertext{2.~update variables $\V{z}$:}
  \V{z}^{(t)}
  &= \argmin_{\V{z}}
  \mathcal{L}_{\rho^{(t)}}\bigl(\V{x}^{(t)}, \V{z}, \V{u}^{(t-1)}\bigr) \notag \\
  &= \argmin_{\V{z}}
  \Bigl\{ \Fdata(\V{z})
  + \frac{\rho^{(t)}}{2} \, \bigNorm{\V{z} - \proxy{\V{z}}^{(t)}}_2^2 \Bigr\}
  \label{eq:z-update}
  \intertext{\hspace*{2.5ex}with:}
  \proxy{\V{z}}^{(t)} &= \V{x}^{(t)} + \V{u}^{(t-1)}/\rho^{(t)} \, ;
  \label{eq:z-proxy-def}
  \intertext{3.~update multipliers $\V{u}$:}
  \V{u}^{(t)} &= \V{u}^{(t-1)} + \rho^{(t)} \,
  \left(\V{x}^{(t)} - \V{z}^{(t)}\right) \, .
  \quad\EndProof
  \label{eq:u-update}
\end{align}

Our algorithm can be seen as an instance of SALSA
\cite{Afonso_et_al-2010-SALSA} with however some improvements.  First, we deal
with the additional constraints that the variables are non-negative. Second,
we allow for changing the weight of the augmented penalty at every iteration
which can considerably speed-up convergence. Third, for real observations, the
operator $\M{H}$ cannot be easily diagonalized (\eg by using FFT) thus the
updating of variables $\V{z}$ cannot be exactly carried out. Finally, we
consider the possibility of warm starting the algorithm with a solution
previously computed.  This latter feature is of interest to improve a solution
if too few iterations have been performed or to find the solution of the
problem with a slightly different value of the hyper-parameter $\mu$.

In Appendix~\ref{sec:proximity-operators}, we show how the updating of
the variables $\V{x}$ (step~1 of Algorithm~1) can be implemented taking into
account the constraints that the parameters are non-negative.  This is our
motivation for imposing $\V{x}\in\Set{X}$ on the variables $\V{x}$ and not on
the variables $\V{z}$.  This avoids introducing additional auxiliary variables
for the sole purpose of accounting for the feasible set.  The formulae to
update the variables $\V{x}\in\Set{X}$ for $\Fsparse(\V{x})$ and
$\Fjoint(\V{x})$ are respectively given by \Eq{eq:sparse-prox-plus} and
\Eq{eq:joint-sparse-prox-plus} in Appendix~\ref{sec:proximity-operators}.

\subsubsection{Solving for the auxiliary variables}
\label{sec:solving-auxiliary-variables}

Since $\mathcal{L}_{\rho}(\V{x},\V{z},\V{u})$ is quadratic with respect to
$\V{z}$, updating these variables (step~2 of Algorithm~1) amounts to solving
the linear problem:
\begin{equation}
  \M{A}^{(t)}\cdot\V{z}^{(t)} = \V{b}^{(t)}
  \label{eq:z-update-linear-problem}
\end{equation}
with:
\begin{align}
  \M{A}^{(t)} &= \M{H}\T\!\!\cdot\M{W}\cdot\M{H} + \rho^{(t)}\,\M{I} \\
  \V{b}^{(t)} &= \M{H}\T\!\!\cdot\M{W}\cdot\V{y} + \rho^{(t)}\,\V{x}^{(t)} + \V{u}^{(t-1)}
\end{align}
with $\M{I}$ the identity matrix (of suitable size).  Since the augmented term
is diagonal and provided data from different spectral channels are
statistically independent, this problem can be solved separately for each
spectral channel:
\begin{equation}
  \M{A}^{(t)}_\ell\cdot\V{z}^{(t)}_\ell = \V{b}_\ell^{(t)} \, ,\quad \forall
  \ell\, ,
\end{equation}
with:
\begin{align}
  \M{A}^{(t)}_\ell &= \M{H}_\ell\T\!\!\cdot\M{W}_\ell\cdot\M{H}_\ell + \rho^{(t)}\,\M{I} \\
  \V{b}^{(t)}_\ell &= \M{H}_\ell\T\!\!\cdot\M{W}_\ell\cdot\V{y}_\ell
  + \rho^{(t)}\,\V{x}_\ell^{(t)} + \V{u}_\ell^{(t-1)}
\end{align}
using the same conventions as in \Eq{eq:direct-model-in-a-channel}.

In practice, we (approximately) solve these problems by means of the conjugate
gradients algorithm \cite{Hestenes_Stiefel-1952-conjugate_gradients,
  Nocedal_Wright-2006-numerical_optimization} and starting with the previous
solution $\V{z}^{(t-1)}$.  Since, in $\M{A}^{(t)}$, the Hessian matrix
$\M{H}\T\!\!\cdot\M{W}\cdot\M{H}$ is regularized by the term
$\rho^{(t)}\,\M{I}$, its condition number is better than that of
$\M{H}\T\!\!\cdot\M{W}\cdot\M{H}$.  We therefore expect that the conjugate
gradients algorithm has a better convergence rate with $\M{A}^{(t)}$ than with
$\M{H}\T\!\!\cdot\M{W}\cdot\M{H}$.  Moreover, according to Eckstein-Bertsekas
theorem \cite{Eckstein_Bertsekas-1992-Douglas-Rachford_splitting}, the ADMM
algorithm is proved to converge provided that approximations in the update of
auxiliary variables $\V{z}$ be absolutely summable.  That is:
\begin{equation}
  \sum_{t = 1}^{\infty} \bigNorm{\V{z}^{(t)} - \V{z}^{(t)}_\Tag{exact}}_2 < \infty
  \label{eq:Eckstein-Bertsekas-constraint}
\end{equation}
must hold with $\V{z}^{(t)}_\Tag{exact}$ the exact solution of
\Eq{eq:z-update} and $\V{z}^{(t)}$ the approximate solution of
\Eq{eq:z-update} returned by the conjugate gradient iterations.  The
demonstration in \cite{Eckstein_Bertsekas-1992-Douglas-Rachford_splitting}
consider ADMM iterations with a fixed quadratic penalty parameter $\rho$, so
it may not strictly apply to our method where $\rho$ is allowed to vary (see
Section~\ref{sec:tuning-rho}).  An easy solution to warrant convergence is to
fix the value of $\rho$ after a certain number of ADMM iterations.
Nevertheless, we observed in our tests that the value of $\rho$ stabilizes to
a fixed value without imposing this.  In
Appendix~\ref{sec:Eckstein-Bertsekas-constraint}, we show how to set the
stopping criterion of the conjugate gradient method so that the constraint in
\Eq{eq:Eckstein-Bertsekas-constraint} holds.  In our tests, we however simply
stop the conjugate gradient iterations when the Euclidean norm of the
residuals of \Eq{eq:z-update-linear-problem} becomes significantly smaller
than its initial value:
\begin{equation}
  \norm{\M{A}^{(t)}\cdot\V{z}^{(t)} - \V{b}^{(t)}}_2 \le \epsilon_\Tag{CG} \,
  \norm{\M{A}^{(t)}\cdot\V{z}^{(t-1)} - \V{b}^{(t)}}_2
\end{equation}
with $\epsilon_\Tag{CG} \in (0,1)$.  For our tests, we took
$\epsilon_\Tag{CG}=10^{-2}$ and allowed a maximum of 5 conjugate gradient
iterations.  With this simple prescription, we did not experiment any
divergence of the global algorithm although it may depend on the problem at
hands and could require to be adapted.

\subsubsection{Stopping criteria}
\label{sec:convergence}

At the solution $\{\V{x}^{*}, \V{z}^{*}, \V{u}^{*}\}$ of problem
(\ref{eq:split-problem}), Karush-Kuhn-Tucker (KKT) conditions of optimality
\cite{Nocedal_Wright-2006-numerical_optimization} stipulate that, the
constraints must be satisfied and that the solution must be a stationary point
of the Lagrangian:
\begin{align}
  &\V{x}^{*} = \V{z}^{*} \label{eq:KKT-1} \\
  &\V{0} \in \partial_{\V{x}} \mathcal{L}_0(\V{x}^{*}, \V{z}^{*}, \V{u}^{*})
  = \mu\,\partial \Fprior(\V{x}^{*}) + \V{u}^{*}
  \label{eq:KKT-2} \\
  &\V{0} \in \partial_{\V{z}} \mathcal{L}_0(\V{x}^{*}, \V{z}^{*}, \V{u}^{*})
  = \partial \Fdata(\V{z}^{*}) - \V{u}^{*}
   \label{eq:KKT-3}
\end{align}
where $\partial$ denotes the subdifferential operator
\cite{Boyd_et_al-2010-method_of_multipliers}. Since $\Fdata$ is
differentiable, $\in$ and $\partial \Fdata$ can be replaced by $=$ and
by $\nabla \Fdata$, the gradient of $\Fdata$ in the third KKT
condition (\ref{eq:KKT-3}) which becomes:
\begin{equation}
  \V{u}^{*} = \nabla \Fdata(\V{z}^{*}) \, .
  \label{eq:KKT-3-alt}
\end{equation}
If $\V{z}^{(t)}$ exactly minimizes $\mathcal{L}_{\rho^{(t)}}(\V{x}^{(t)},
\V{z}, \V{u}^{(t-1)})$, we have:
\begin{align}
  &\nabla \Fdata(\V{z}^{(t)}) - \V{u}^{(t-1)}
  + \rho^{(t)} \, (\V{z}^{(t)} - \V{x}^{(t)}) = \V{0} \notag \\
  \Longrightarrow\
  &\nabla \Fdata(\V{z}^{(t)}) = \V{u}^{(t)} \, ,
  \label{eq:KKT-3-admm}
\end{align}
thus the 3rd KKT condition in \Eq{eq:KKT-3-alt} is automatically satisfied at
the end of an \emph{exact} ADMM iteration.

As we use the method of conjugate gradients to solve
\Eq{eq:z-update-linear-problem}, \Eq{eq:KKT-3-admm} is only approximately
satisfied.  Moreover, updating the multipliers $\V{u}$ according to step~3 of
the algorithm may be subject to accumulation of rounding errors.  The
stability of the algorithm or its convergence rate may be improved by taking
$\V{u}^{(t)} = \nabla \Fdata(\V{z}^{(t)})$.  In our tests, tough, we have not seen
significant differences between updating the Lagrange multipliers according to
\Eq{eq:u-update} or according to \Eq{eq:KKT-3-admm}.

Since $\V{x}^{(t)}$ minimizes $\mathcal{L}_{\rho^{(t)}}(\V{x},
\V{z}^{(t-1)}, \V{u}^{(t-1)})$, we have:
\begin{align}
  \V{0}
  &\in \mu \, \partial \Fprior(\V{x}^{(t)}) + \V{u}^{(t-1)}
  + \rho^{(t)} \, (\V{x}^{(t)} - \V{z}^{(t-1)}) \notag \\
  &\in \mu \, \partial \Fprior(\V{x}^{(t)}) + \V{u}^{(t)}
  + \rho^{(t)} \, (\V{z}^{(t)} - \V{z}^{(t-1)}) \notag \\
  \Longrightarrow&\quad
  - \rho^{(t)} \, (\V{z}^{(t)} - \V{z}^{(t-1)})
  \in \mu \, \partial \Fprior(\V{x}^{(t)}) + \V{u}^{(t)}
  \notag
\end{align}
thus:
\begin{equation}
  \V{s}^{(t)} = \rho^{(t)} \, \bigl(\V{z}^{(t)} - \V{z}^{(t - 1)}\bigr)
  \label{eq:dual-residual}
\end{equation}
can be seen as the residuals for the 2nd KKT condition in \Eq{eq:KKT-2},
while:
\begin{equation}
  \V{r}^{(t)} = \V{x}^{(t)} - \V{z}^{(t)}
  \label{eq:primal-residual}
\end{equation}
are the residuals for the primary constraint in \Eq{eq:KKT-1}.

Finally, the KKT conditions imply that the so-called \emph{primal} and
\emph{dual residuals} \cite{Boyd_et_al-2010-method_of_multipliers} defined in
\Eq{eq:primal-residual} and \Eq{eq:dual-residual} must converge to zero.
Following \cite{Boyd_et_al-2010-method_of_multipliers}, we therefore stop the
algorithm when:
\begin{equation}
  \bigl\Vert \V{r}^{(t)} \bigr\Vert_2 \le \tau_\Tag{prim}^{(t)}
  \quad\text{and}\quad
  \bigl\Vert \V{s}^{(t)} \bigr\Vert_2 \le \tau_\Tag{dual}^{(t)} \, ,
  \label{eq:convergence-criterion}
\end{equation}
where the convergence thresholds are given by:
\begin{align}
  \tau_\Tag{prim}^{(t)} &\bydef \sqrt{N}\,\epsilon_\Tag{abs} + \epsilon_\Tag{rel}
  \, \max\bigl(\bigl\Vert \V{x}^{(t)} \bigr\Vert_2,
               \bigl\Vert \V{z}^{(t)} \bigr\Vert_2 \bigr) \, ,
  \label{eq:primal-threshold} \\
  \tau_\Tag{dual}^{(t)} &\bydef \sqrt{N}\,\epsilon_\Tag{abs} +
  \epsilon_\Tag{rel} \, \bigl\Vert \V{u}^{(t)} \bigr\Vert_2 \, ,
  \label{eq:dual-threshold}
\end{align}
where $N = \Card(\V{x})$ is the number of sought parameters,
$\epsilon_\Tag{abs} \ge 0$ and $\epsilon_\Tag{rel} \in (0,1)$ are absolute and
relative convergence tolerances.  For our tests, we found that
$\epsilon_\Tag{abs} = 0$ and $\epsilon_\Tag{rel} = 10^{-3}$ yield sufficient
precision for the solution.

\subsubsection{Initialization and warm start}
\label{sec:warm-start}

Given an initial estimate $\V{z}^{(0)}$ for the auxiliary variables, the 3rd
KKT condition in \Eq{eq:KKT-3-alt} suggests to start the iterative algorithm
with initial Lagrange multipliers given by:
\begin{equation}
  \V{u}^{(0)} = \nabla \Fdata\bigl(\V{z}^{(0)}\bigr)
  = \M{H}\T\!\!\cdot\M{W}\cdot\bigl(\M{H}\T\!\!\cdot\V{z}^{(0)} - \V{y}\bigr) \, .
\end{equation}
Starting the algorithm as suggested has several advantages.  First, there are
no needs for the initial variables $\V{z}^{(0)}$ to belong to the feasible
set.  Second, this readily provides initial Lagrange multipliers.  Note that
starting with an initial estimate $\V{x}^{(0)}$ for the variables $\V{x}$
would have the double drawback that $\V{x}^{(0)}$ must be feasible and that
since $\Fprior(\V{x})$ may be non-differentiable it does not yields an
explicit expression for the initial Lagrange multipliers.

The other required initial setting is the value of the augmented penalty
parameter $\rho^{(1)}$ used to compute the first estimate $\V{x}^{(1)}$ of the
variables $\V{x}$ given $\V{z}^{(0)}$ and $\V{u}^{(0)}$.  For the subsequent
iterations, $\rho$ can be kept constant or updated according to the
prescription described in Sec.~\ref{sec:tuning-rho}.

To continue the iterations or compute a solution with slightly different
parameters (\eg the regularization parameter $\mu$), the possibility to
restart the algorithm with the output of a previous run with no loss of
performances regarding the rate of convergence is a needed feature.  This is
called \emph{warm restart} and is simply achieved by saving a minimal set of
variables upon return of the algorithm.  Since each iteration of our algorithm
starts by computing the variables $\V{x}$ given the auxiliary variables
$\V{z}$, the Lagrange multipliers $\V{u}$ and the augmented penalty parameter
$\rho$, it is sufficient to save $\{\V{z},\V{u},\rho\}$ for being able to warm
restart the method.

\subsubsection{Tuning the augmented penalty parameter $\rho$}
\label{sec:tuning-rho}

One of the important settings of the ADMM method is the value of the augmented
penalty parameter $\rho$: if it is too small, the primal constraints $\V{x} =
\V{z}$ will converge too slowly; while the cost functions will decrease too
slowly if $\rho$ is too large.  Some authors, \eg
\cite{Afonso_et_al-2010-SALSA}, use a constant augmented penalty parameter for
all the iterations which requires trials and errors to find an efficient value
for $\rho$.  In fact, it is worth using a good value of $\rho$ at every
iteration of ADMM to accelerate the convergence
\cite{Boyd_et_al-2010-method_of_multipliers}.  In this section, we describe
means to automatically derive a suitable value for the augmented penalty
parameter following a simple reasoning.

The convergence criterion defined in \Eq{eq:convergence-criterion} is
equivalent to have:
\begin{equation}
  \phi^{(t)} \le 1
  \quad\text{with}\quad
  \phi^{(t)} \bydef \max\left(
  \frac{\bigl\Vert \V{r}^{(t)} \bigr\Vert_2}{\tau_\Tag{prim}^{(t)}},
  \frac{\bigl\Vert \V{s}^{(t)} \bigr\Vert_2}{\tau_\Tag{dual}^{(t)}}
  \right) .
  \label{eq:convergence-criterion-alt}
\end{equation}
According to the updating rules in one ADMM iteration (see Algorithm~1),
$\phi^{(t)}$ does only depend on $\V{z}^{(t-1)}$, $\V{u}^{(t-1)}$ and
$\rho^{(t)}$.  The augmented penalty parameter $\rho^{(t)}$ is therefore the
only tunable parameter that has an incidence on the value of $\phi^{(t)}$ for
the $t$\textsup{th} ADMM iteration. The idea is then to chose the value of
$\rho^{(t)}$ so as to approximately minimize $\phi^{(t)}$.  In terms of number
of ADMM iterations, we expect to achieve the faster convergence of the
algorithm in that way.  However, tuning $\rho$ at every ADMM iteration
requires to repeat each iteration for different values of $\rho$ and has
therefore the same computational cost as several ADMM iterations.  A
compromise has to be found between the accuracy on $\rho$ and the number of
trials.

Our objective is to derive an economical way to find:
\begin{equation}
  \rho^{(t)} \approx \rho^{(t)}_{*} \bydef \argmin_{\rho} \phi_{t}(\rho) \, .
  \label{eq:best-rho}
\end{equation}
where $\phi_{t}(\rho)$ is the value taken by $\phi^{(t)}$ when $\rho^{(t)} =
\rho$.  All the quantities, $\norm{\V{r}^{(t)}}_2$, $\norm{\V{s}^{(t)}}_2$,
$\tau_\Tag{prim}^{(t)}$, and $\tau_\Tag{dual}^{(t)}$, involved in $\phi^{(t)}$
vary continuously (though not necessarily smoothly) with respect to
$\rho^{(t)}$; hence, considering the definition of $\rho^{(t)}_{*}$ and
$\phi^{(t)}$, we obtain the following implication:
\begin{equation}
  \rho^{(t)} = \rho^{(t)}_{*}
  \Longrightarrow
  \frac{\bigl\Vert \V{r}^{(t)} \bigr\Vert_2}{\tau_\Tag{prim}^{(t)}} =
  \frac{\bigl\Vert \V{s}^{(t)} \bigr\Vert_2}{\tau_\Tag{dual}^{(t)}} \, .
  \label{eq:best-rho-necessary-condition}
\end{equation}
Besides, the norm of the primal residuals $\norm{\V{r}^{(t)}}_2$ is a
decreasing function of $\rho^{(t)}$; while the norm of the dual residuals
$\norm{\V{s}^{(t)}}_2$ is an increasing function of $\rho^{(t)}$
\cite{Boyd_et_al-2010-method_of_multipliers} and, close enough to the
solution, the values of $\tau_\Tag{prim}$ and $\tau_\Tag{dual}$ should
converge to their final values and thus not depend too much on $\rho$.  Under
these assumptions, the ratios $\norm{\V{r}^{(t)}}_2/\tau_\Tag{prim}^{(t)}$ and
$\norm{\V{s}^{(t)}}_2/\tau_\Tag{dual}^{(t)}$ should also be decreasing and
increasing functions of $\rho$ respectively.  Close to the solution
$\{\V{x}^{*}, \V{z}^{*}, \V{u}^{*}\}$ of the problem, the necessary condition
in \Eq{eq:best-rho-necessary-condition} is therefore also a sufficient
condition to define the optimal value $\rho^{(t)}_{*}$.  These considerations
lead us to choose $\rho^{(t)}$ such that:
\begin{equation}
  \eta^{(t)} \approx 1
  \quad\text{with}\quad
  \eta^{(t)} \bydef
  \frac{\bigl\Vert \V{r}^{(t)} \bigr\Vert_2 \, \tau_\Tag{dual}^{(t)}}
       {\bigl\Vert \V{s}^{(t)} \bigr\Vert_2 \, \tau_\Tag{prim}^{(t)}} \, ,
  \label{eq:eta}
\end{equation}
which is expected to be a decreasing function of $\rho^{(t)}$ close to the
solution.  A better alternative may be to choose $\rho^{(t)}$ such that:
\begin{equation}
  \eta^{(t)}_\Tag{alt} \approx 1
  \quad\text{with}\quad
  \eta^{(t)}_\Tag{alt} \bydef
  \frac{\bigl\Vert \V{r}^{(t)} \bigr\Vert_2 \, \tau_\Tag{dual}^{(t - 1)}}
       {\bigl\Vert \V{s}^{(t)} \bigr\Vert_2 \, \tau_\Tag{prim}^{(t - 1)}} \, .
  \label{eq:eta-alt}
\end{equation}
Indeed, as $\tau_\Tag{dual}^{(t - 1)}$ and $\tau_\Tag{prim}^{(t - 1)}$ do not
depend on $\rho^{(t)}$, $\eta^{(t)}_\Tag{alt}$ is always decreasing function
of $\rho^{(t)}$ while approaching $\eta^{(t)}$ when the algorithm is close to
the solution.  The following algorithm implements our safeguarded strategy to
find $\rho^{(t)}>0$ such that $\eta^{(t)} \approx 1$ or $\eta^{(t)}_\Tag{alt}
\approx 1$.

\smallskip

\textbf{Algorithm 2.}  \emph{Tuning of the augmented penalty parameter
  $\rho$ so that $\eta \approx 1$.}  Choose $\sigma \ge 0$, $\tau > 1$, $\gamma
> 1$ and an initial value for $\rho$, and set $\rho_\Tag{min} = 0$,
$\rho_\Tag{max} = +\infty$.  Then, until convergence, repeat the following
steps:
\begin{enumerate}
\item Update $\V{x}$, $\V{z}$, and $\V{u}$ according to ADMM updating rules.
  Compute $\eta$, defined in \Eq{eq:eta} or in \Eq{eq:eta-alt}, and $\phi$,
  defined in \Eq{eq:convergence-criterion-alt}.
\item If $1/\tau \le \eta \le \tau$ or $\phi < \sigma\,\phi^{(t-1)}$, accept
  the current solution and stop.
\item If $\eta < 1/\tau$, then $\rho$ is too large; let $\rho_\Tag{max} :=
  \rho$ and
  \begin{displaymath}
    \rho := \left\{
    \begin{array}{ll}
      \sqrt{\rho_\Tag{min}\,\rho_\Tag{max}} & \text{if } \rho_\Tag{min} > 0\\
      \rho_\Tag{max}/\gamma & \text{otherwise}\\
    \end{array}\right.
  \end{displaymath}
  then go to step 1.
\item If $\eta > \tau$, then $\rho$ is too small; let
  $\rho_\Tag{min} := \rho$ and
  \begin{displaymath}
    \rho := \left\{
    \begin{array}{ll}
      \sqrt{\rho_\Tag{min}\,\rho_\Tag{max}} & \text{if } \rho_\Tag{max} < \infty\\
      \gamma\,\rho_\Tag{min} & \text{otherwise}\\
    \end{array}\right.
  \end{displaymath}
  then go to step 1. \EndProof
\end{enumerate}

The following remarks clarify some aspects of this algorithm:
\begin{itemize}
\item To simplify the notations, we dropped the index $t$ of the ADMM
  iteration in the equations of Algorithm~2.  The updating of variables
  (step~1) must be understood as computing $\V{x}^{(t)}$, $\V{z}^{(t)}$, \etc
  given $\V{z}^{(t-1)}$ and $\V{u}^{(t-1)}$, and assuming $\rho^{(t)} = \rho$.
\item Except for the very first ADMM iteration ($t=1$), the initial value for
  $\rho$ is the previous selected value $\rho^{(t-1)}$.  For the first iteration,
  we derive an initial value of $\rho$ such that:
  \begin{align}
    \rho^{(1)}
    & = \argmin_{\rho} \Fdata\!\left(\V{z}^{(1)} - \V{u}^{(1)}/\rho\right)
    \notag\\
    & = \frac{
      \V{u}^{(1)\TransposeLetter}\!\!\cdot
      \M{H}\T\!\!\cdot\M{W}\cdot\M{H}\cdot
      \V{u}^{(1)}
    }{
      \V{u}^{(1)\TransposeLetter}\!\!\cdot
      \V{u}^{(1)}
    }
  \end{align}
  with $\V{u}^{(1)} = \nabla\Fdata(\V{z}^{(1)}) =
  \M{H}\T\!\!\cdot\M{W}\cdot(\M{H}\cdot\V{z}^{(1)} - \V{y})$ and which amounts
  to have the best $\proxy{\V{x}}^{(1)}$, defined in \Eq{eq:x-proxy-def}, with
  respect to $\Fdata$.  This choice has the advantage of avoiding an
  initialization with an arbitrary value for $\rho$.  The rule does however
  not yield an efficient strategy for tuning $\rho$ at every iteration.
\item Algorithm~2 is safeguarded in the sense that it maintains a strict
  bracketing $\rho_\Tag{min} < \rho^{(t)}_{*} < \rho_\Tag{max}$ of the
  solution.
\item In step~2 of Algorithm~2: The value of $\rho$ is accepted when $1/\tau
  \le \eta \le \tau$ which, with $\tau > 1$, is how we express that $\eta
  \approx 1$.  We achieved good results with $\tau=1.2$ in our tests.  The
  current value of $\rho$ is also accepted, if the relative reduction in the
  convergence criterion $\phi^{(t)}$, defined in
  \Eq{eq:convergence-criterion-alt}, is better than $\sigma$ with respect to
  the previous iteration.  This shortcut helps to reduce the number of inner
  iterations.  To avoid this shortcut, it is sufficient to take: $\sigma=0$.
  We took $\sigma=0.9$ in our tests.
\item In step~3 and step~4 of Algorithm~2: When $\rho$ has been bracketed by
  $(\rho_\Tag{min},\rho_\Tag{max})$, taking $\rho =
  \sqrt{\rho_\Tag{min}\,\rho_\Tag{max}}$, that is the geometrical means of the
  end points, is similar to a bisection step in a zero finding algorithm.
\item For the first ADMM iteration, the magnitude of $\rho$ is not yet known
  so to avoid too many iterations, we use a larger value of the loop gain
  $\gamma$, say $\gamma = 10$ when $t=1$ and $\gamma = 1.5$ for $t>1$.
\end{itemize}

Another possibility is to always accept an ADMM iteration and simply use the
value of $\eta^{(t)}$ to determine whether $\rho$ should be reduced, kept the
same, or augmented for the next iteration.  For instance:
\begin{equation}
  \rho^{(t+1)} = \left\{
  \begin{array}{ll}
    \gamma \, \rho^{(t)} & \text{if $\eta_t > \tau$,} \\
    \rho^{(t)} / \gamma  & \text{if $\eta_t < 1/\tau$,} \\
    \rho^{(t)}           & \text{else;} \\
  \end{array}
  \right.
  \label{eq:rho-forward-update}
\end{equation}
with $\tau \ge 1$ and $\gamma > 1$.  In words, $\rho$ is augmented (by
multiplying it by a factor $\gamma$) whenever the relative size of the primal
residuals is significantly larger than that of the dual residuals; while
$\rho$ is reduced (by dividing it by a factor $\gamma$) whenever the relative
size of the primal residuals is significantly smaller than that of the dual
residuals.  This strategy is similar to the one described by Boyd \emph{et
  al.}  \cite{Boyd_et_al-2010-method_of_multipliers} except that our
prescription properly scales with the magnitudes of the residuals and of the
objective function of the problem so we expect a better behavior.

%
%

\subsection{Debiasing the solution}
\label{sec:debiasing}

\begin{figure}
  \centering
  \includegraphics[width=42mm]{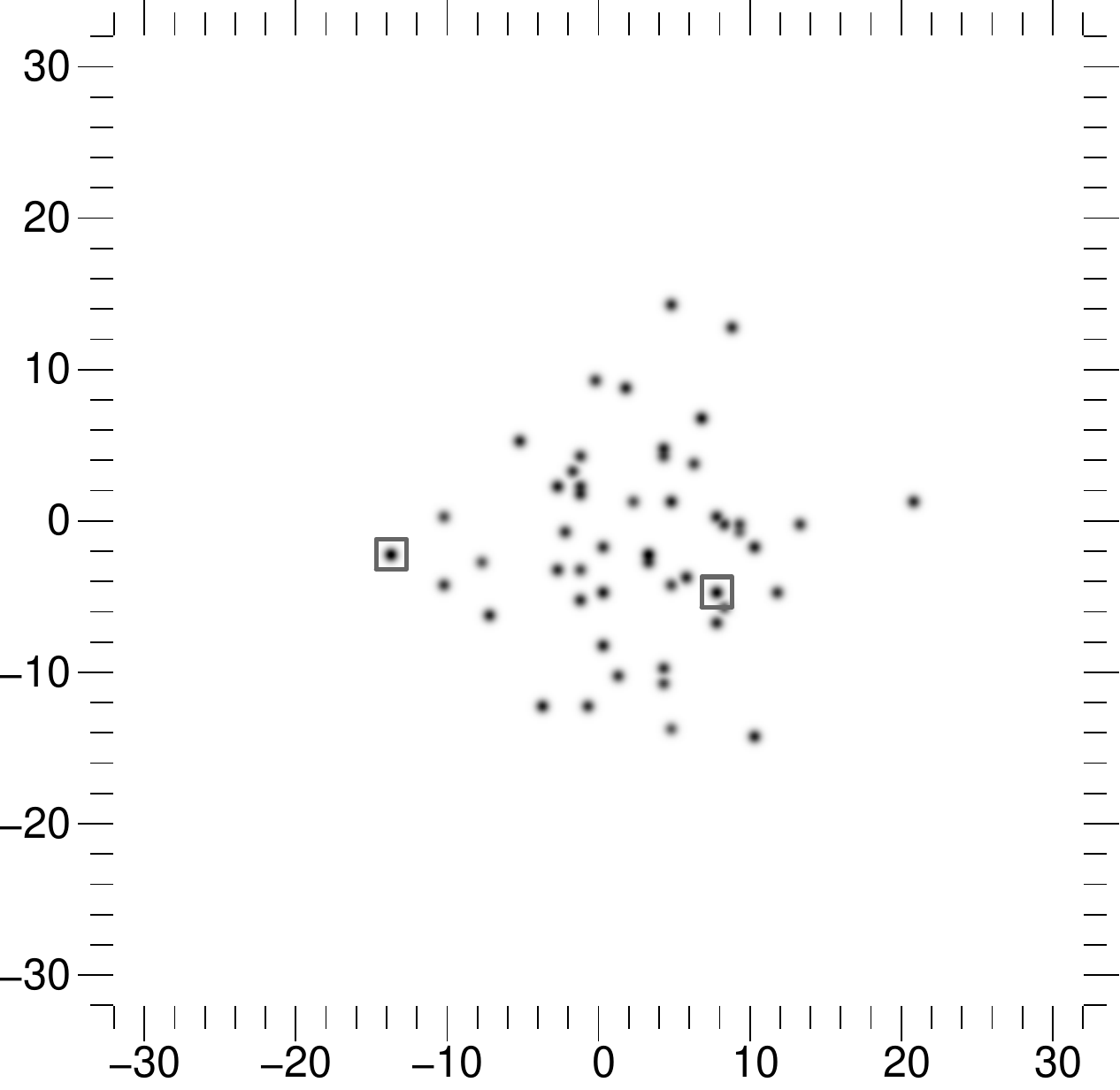}
  \includegraphics[width=42mm]{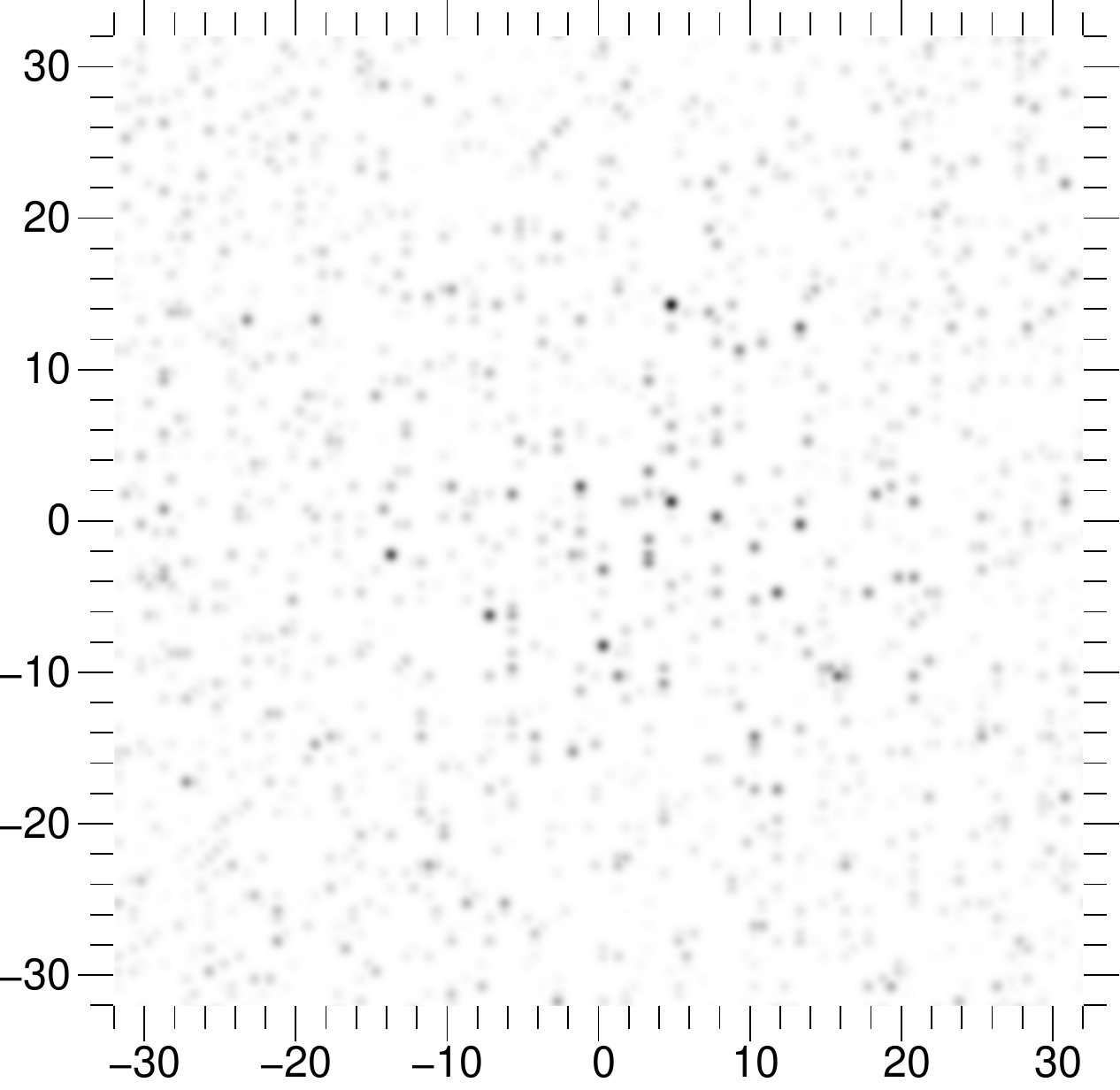}\\
  \includegraphics[width=42mm]{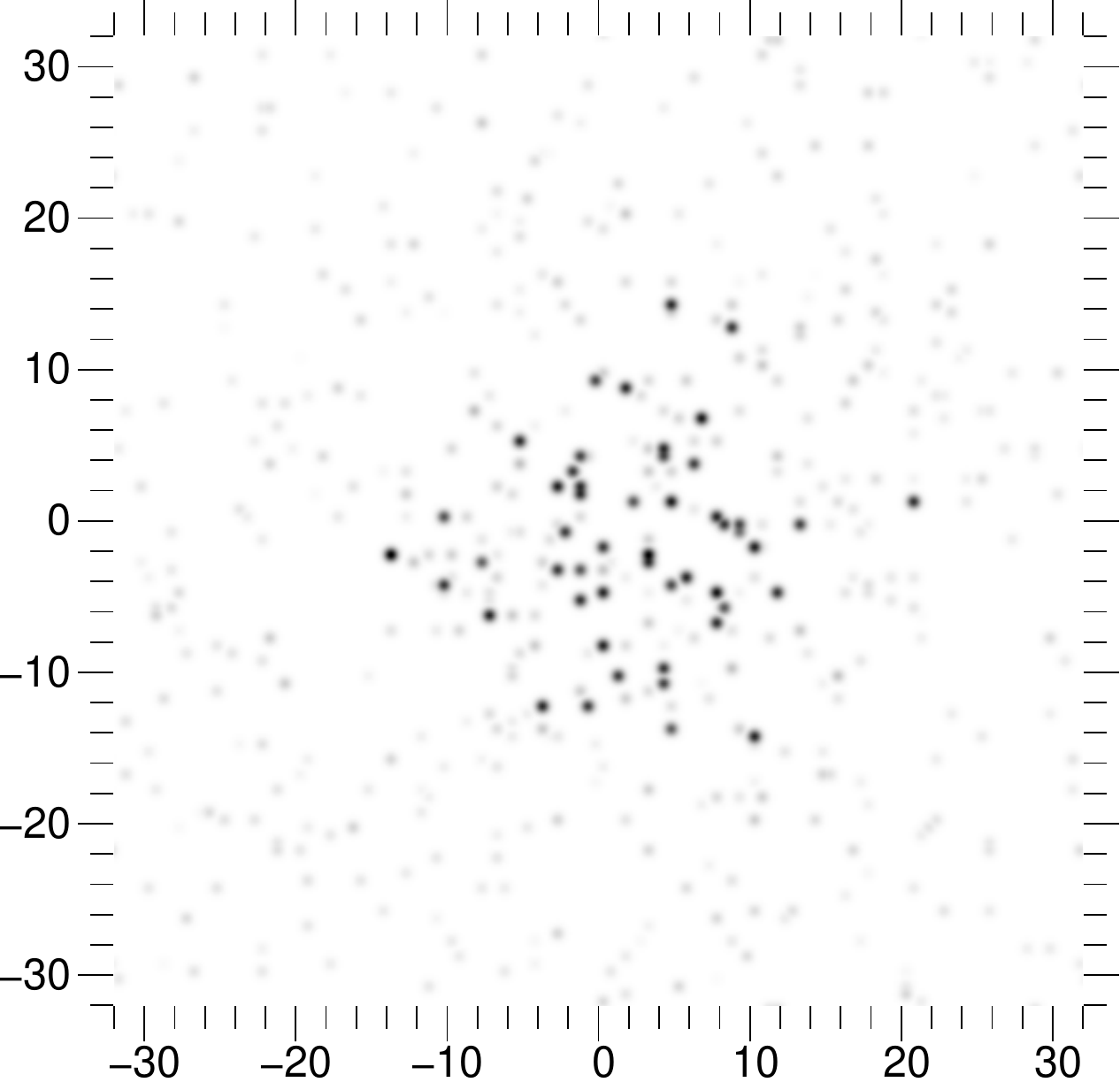}
  \includegraphics[width=42mm]{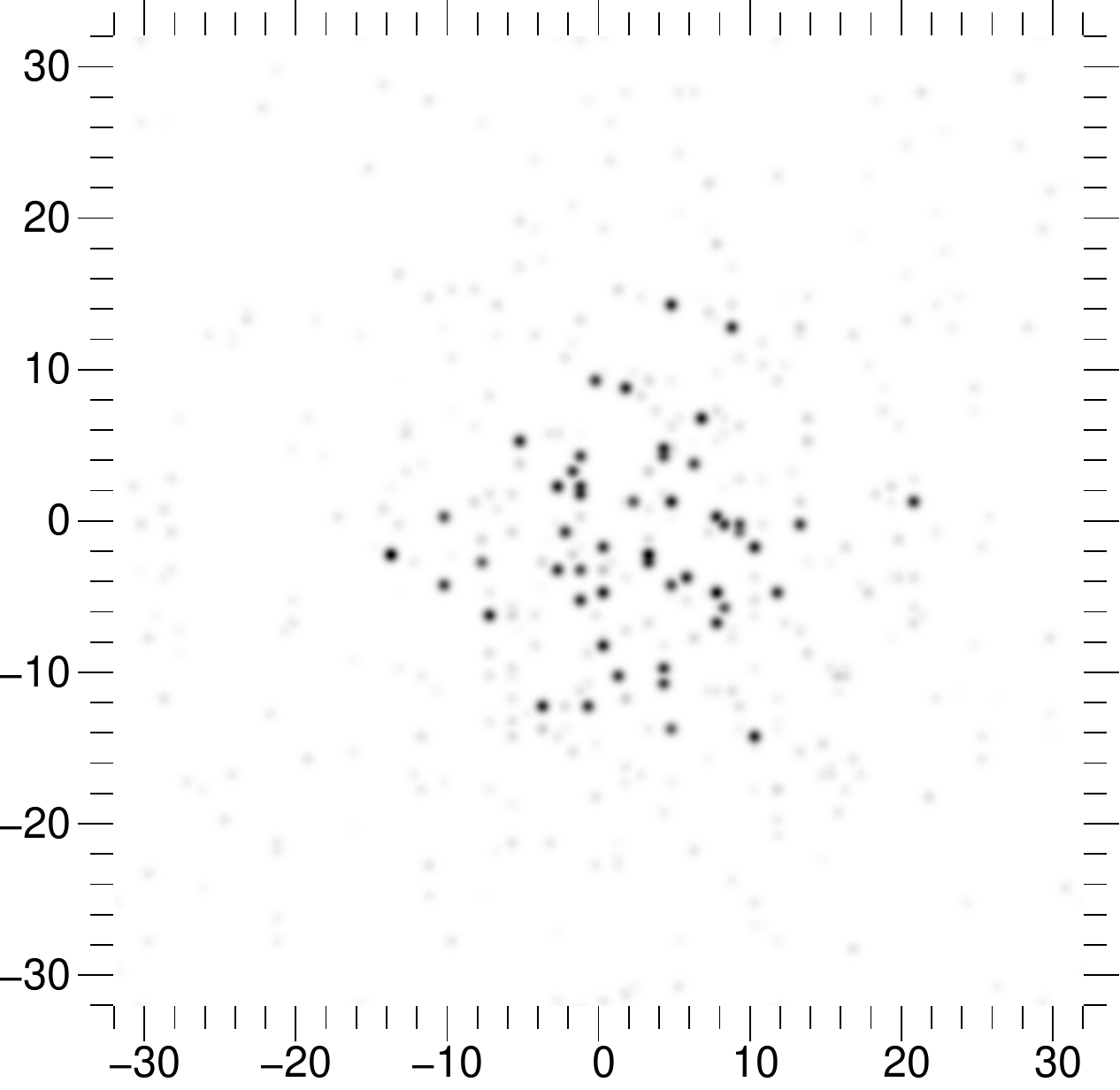}
  \caption{Integrated flux for the star cluster. From top to bottom, left to
    right: true object; reconstruction with fully separable sparsity prior
    $\Fsparse(\V{x})$ defined in Eq.~(\ref{eq:sparse-prior}); reconstruction
    assuming a gray object and with spatial sparsity $\Fsparse(\V{g})$ defined
    in Eq.~(\ref{eq:gray-regularization}); reconstruction with joint-sparsity
    prior $\Fjoint(\V{x})$ defined in Eq.~(\ref{eq:joint-sparse-prior}).  The
    spectra of the sources encircled by the boxes are shown in
    Fig.~\ref{fig:spectra}.  Axes units are in
    milliarcseconds. \label{fig:integrated-flux}}
\end{figure}

\begin{figure}
  \centering
  \includegraphics[width=80mm]{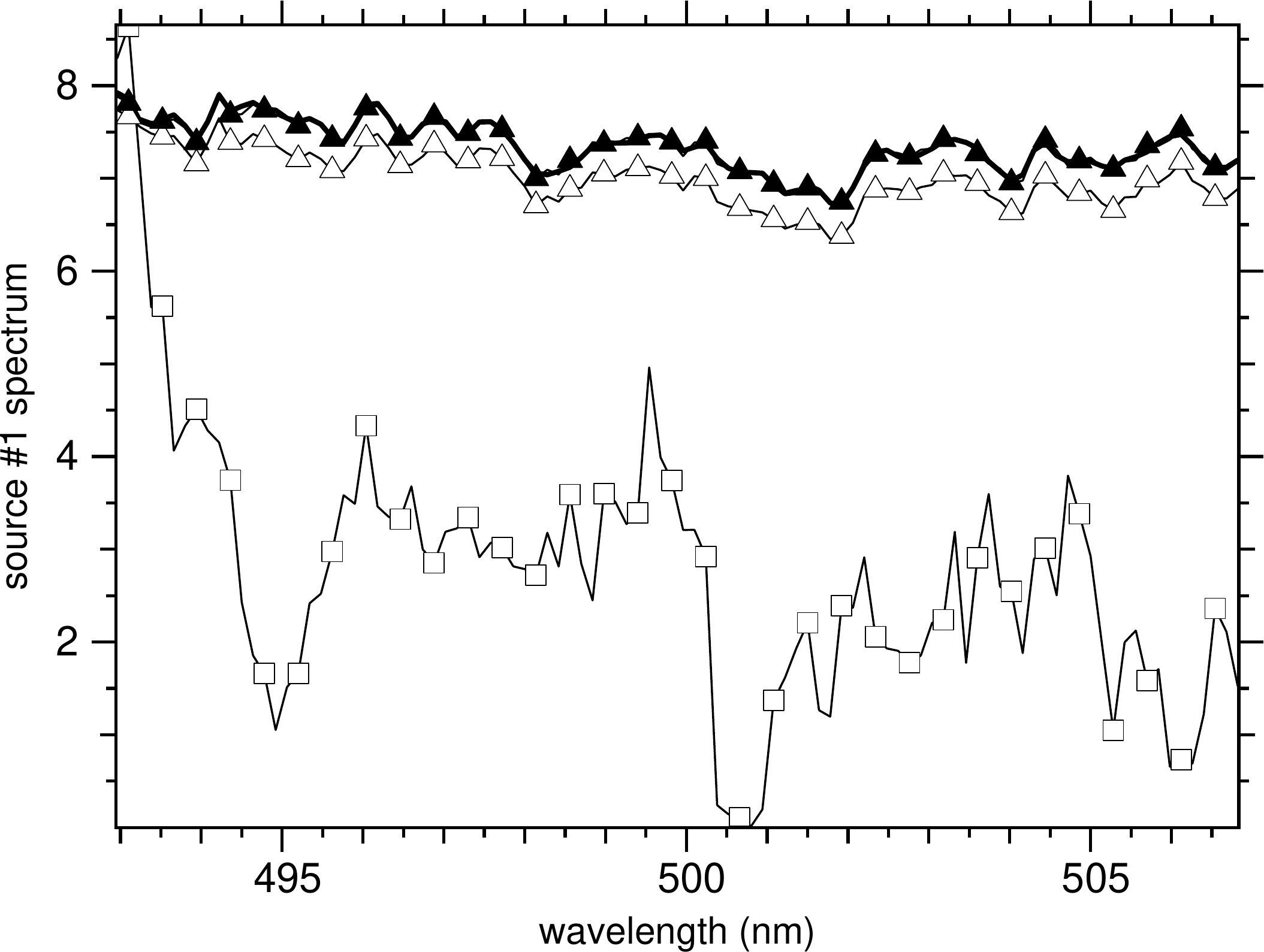}
  \includegraphics[width=80mm]{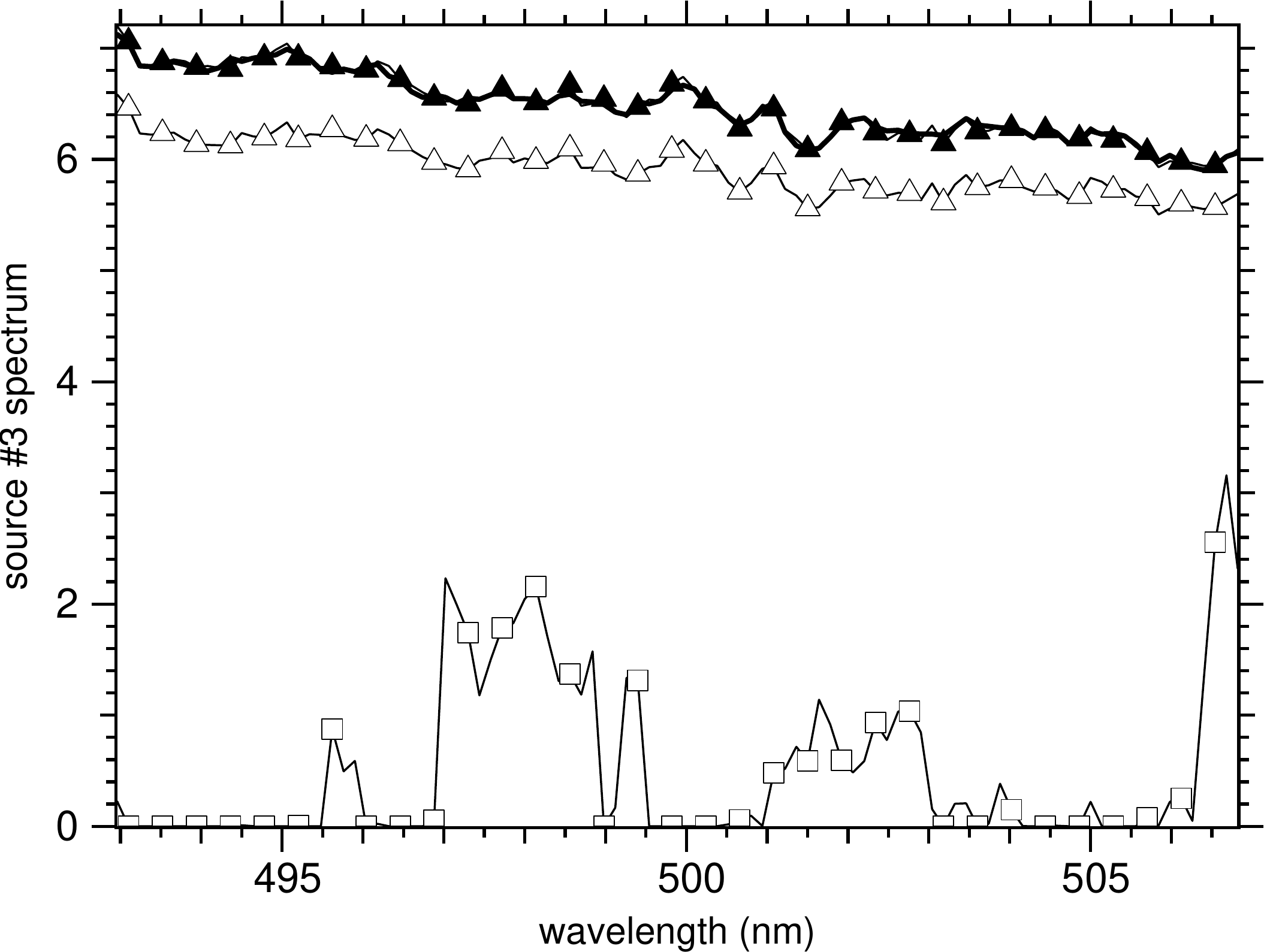}
  \caption{Spectra of two selected sources.  Each panel show the spectra of
    one of the sources encircled by the boxes in
    Fig.~\ref{fig:integrated-flux}. Thick lines are for the true spectra and
    thin lines with markers for the restored spectra.  The open squares
    indicate the reconstruction with fully separable sparsity prior
    $\Fsparse(\V{x})$; the open triangles indicate the reconstruction with
    joint sparsity prior $\Fjoint(\V{x})$; the filled triangles indicate the
    restored spectra after debiasing (which are virtually indistinguishable
    from the true ones).  \label{fig:spectra}}
\end{figure}

One of the drawback of sparsity imposed by means of the $\ell_1$ norm is that
it yields a result which is biased toward zero
\cite{Figueiredo_et_al-2007-gradient_projection}.  For the simulations
presented in Fig.~\ref{fig:integrated-flux}, this bias can be seen on the
recovered spectra in Fig.~\ref{fig:spectra} and in the brightness
distributions of the pixels in Fig.~\ref{fig:mean-flux-histo}.  Since the
sparsity constraint really improves the detection of the sources, the
resulting image can be used to decide where the sources are.  By thresholding
the gray image or the wavelength integrated multi-spectral image resulting
from a reconstruction with a spatial sparsity constraint, we define a sparse
spatial support $\mathcal{S}$ containing all detected sources.  Then, as
proposed in \cite{Wright_et_al-2009-SPARSA}, for the debiasing step, we
minimize the likelihood function $\Fdata(\V{x})$ with non-negative constraints
only over $\V{x}_\mathcal{S}$ defined as the parameters $\V{x}$ restricted to
the support $\mathcal{S}$ while keeping all other parameters equal to zero.
As the sub-matrix $\M{H}_\mathcal{S}$ containing the columns of $\M{H}$
restricted by $\mathcal{S}$ is well conditionned, no additionnal prior is
needed to define the debiased solution.


\section{Results}

To check the proposed algorithm, we simulated a cluster of 50 stars with
random positions and luminosities and with spectra randomly taken from the
library compiled by Jacoby et al. \cite{Jacoby_et_al-1984-stellar_spectra}.
The field of view is $128\times128$ pixels with
$0.5\,\mathrm{milliarcseconds/pixel}$ and we took $100$ spectral channels from
$\lambda = 493\,\mathrm{nm}$ to $\lambda = 507 \,\mathrm{nm}$ by steps of
$\Delta\lambda=0.14\,\mathrm{nm}$. To simulate the observations, we took $100$
random interferometric baselines with a maximum baseline of $180\,\mathrm{m}$.
We added Gaussian white noise to the complex visibilities with a level such
that the maximum signal-to-noise ratio (SNR) is equal to 100.

For the image reconstructions, we considered three different cases: the
reconstruction of a multi-spectral distribution with the regularization
$\Fsparse(\V{x})$ in \Eq{eq:sparse-prior}, or the regularization
$\Fjoint(\V{x})$ in \Eq{eq:joint-sparse-prior}, and the reconstruction
of a gray object $\V{g}$ with the regularization $\Fsparse(\V{g})$.  In
order to set the relative weight of the priors, we choose the value of the
hyper-parameter $\mu$ which yields an image which has the least mean square
error with the true distribution.  Once the values of $\mu$ and $\rho$ are
chosen, the reconstruction of a $128\times128\times100$ distribution from
$\sim2\times10^4$ measurements takes about 4 minutes on a GNU/Linux
workstation with a quad-core processor at $3\,\mathrm{GHz}$ and using a
multi-threaded version of FFTW \cite{FFTW3} to compute the discrete Fourier
transforms.

\begin{figure}
  \centering
  \includegraphics[width=80mm]{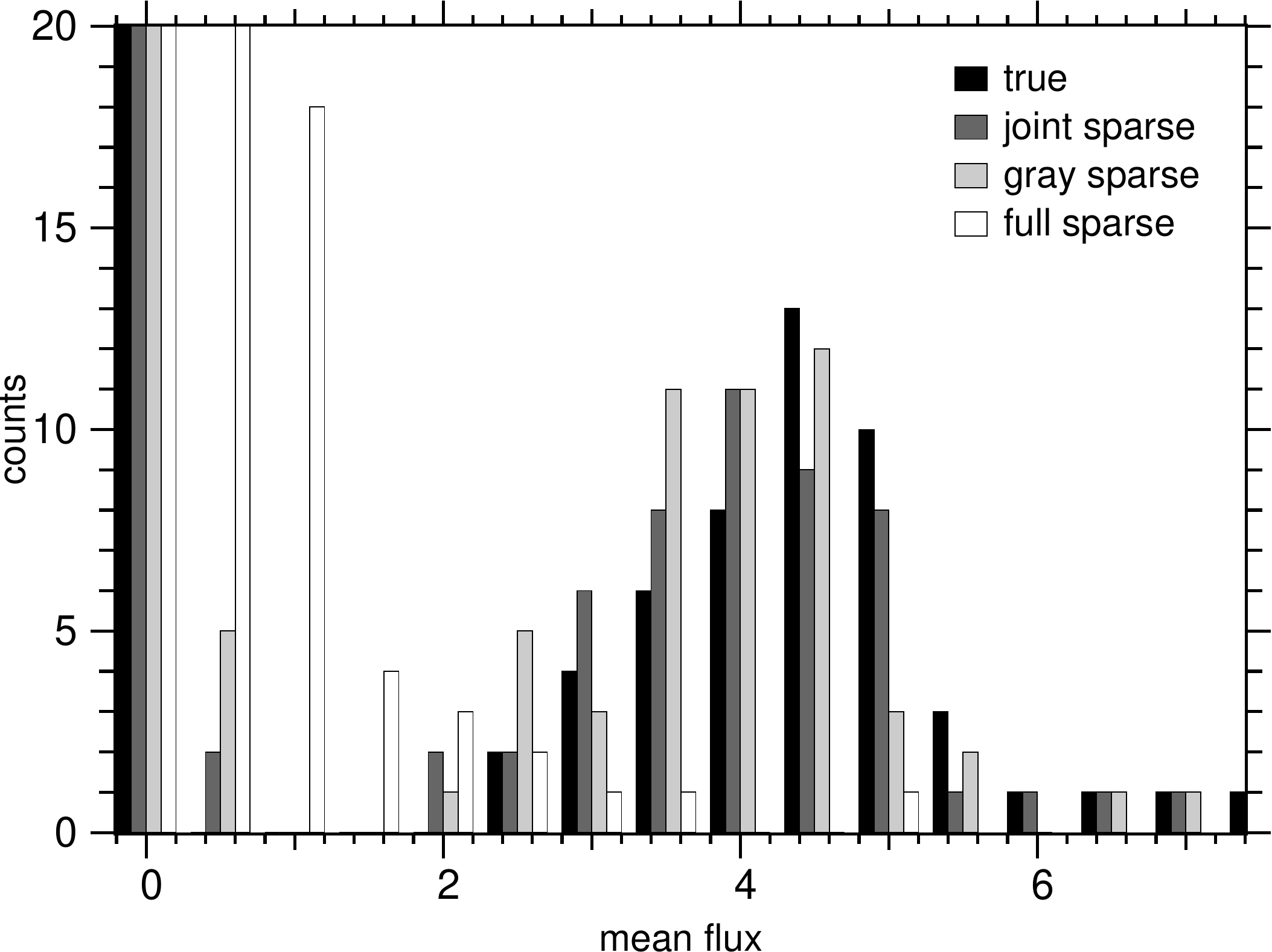}
  \caption{Histograms of the mean fluxes of the sources for the true object
    (in \emph{black}), for the 3-D images restored with fully separable
    sparsity (in \emph{white}) and with joint-sparsity (in \emph{dark gray})
    priors, and for the 2-D gray image restored with sparsity prior (in
    \emph{light gray}).  The vertical scale has been truncated to focus on the
    distributions of the brightest sources. \label{fig:mean-flux-histo}}
\end{figure}

\begin{figure}
  \centering
  \includegraphics[width=80mm]{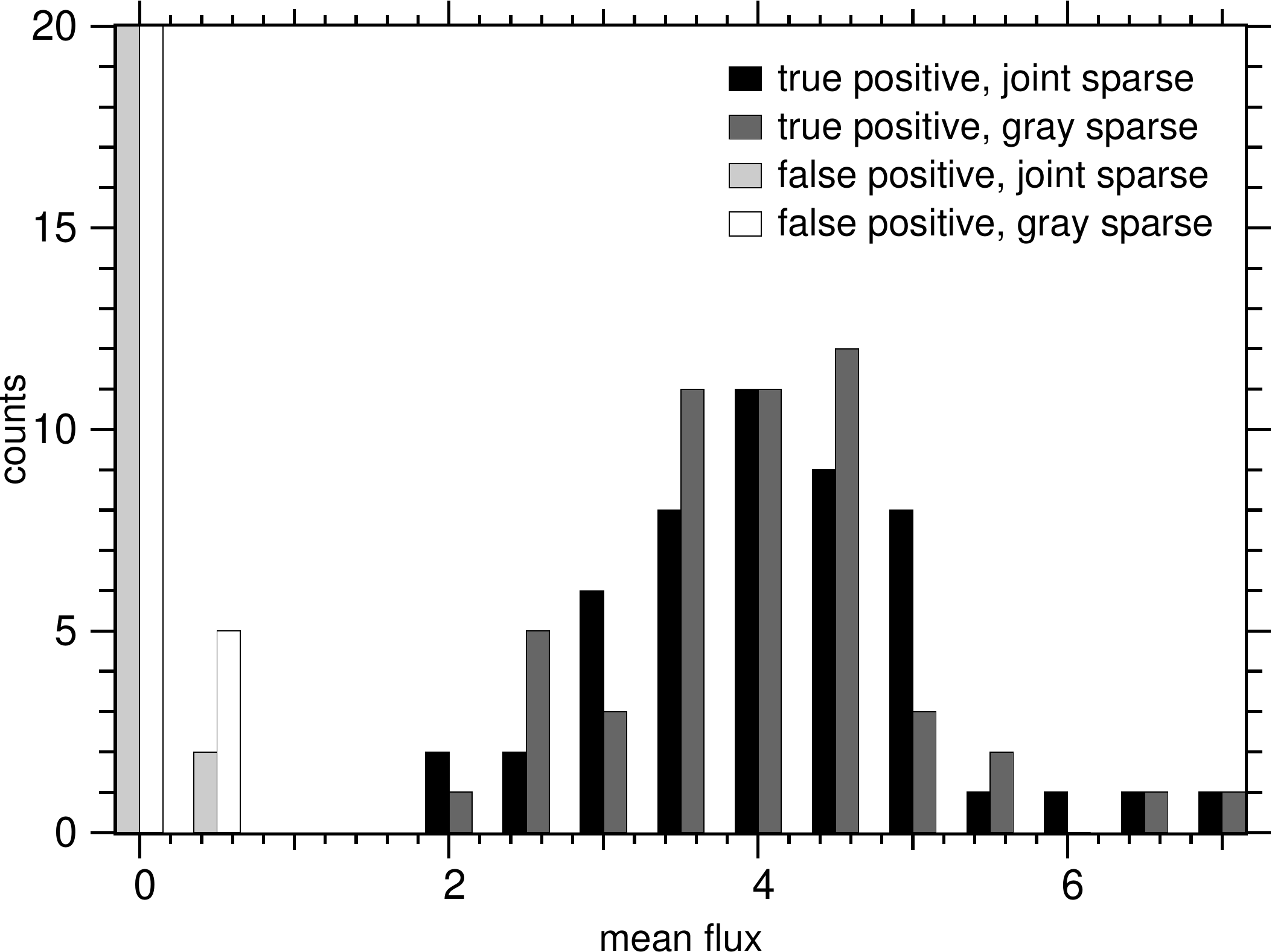}
  \caption{Histograms of the mean flux of the true and false positive
    detection in the reconstructions under joint sparsity and gray sparsity
    priors.  A positive detection is defined as a pixel with non-zero mean
    flux in the reconstruction.  The vertical scale has been truncated to
    focus on the distributions of the true positive
    sources. \label{fig:detect}}
\end{figure}

Figure~\ref{fig:integrated-flux} shows the integrated flux, \ie $\sum_\ell
x_{n,\ell}$, for the true distribution and for the reconstructed ones.  In all
cases, sparsity priors effectively yield a solution with point-like
structures.  However, when there is no trans-spectral constraints, only a few
sources are correctly found and there are many more spurious sources.  When
using $\Fjoint(\V{x})$ or assuming a gray object, the estimated integrated
luminosity is much more consistent with that of the true object: all existing
sources are found and the spurious sources are not only less numerous but also
much fainter than the true ones.  This is shown by the brightness
distributions depicted by Fig.~\ref{fig:mean-flux-histo} and
Fig.~\ref{fig:detect}.

Figure~\ref{fig:spectra} shows the spectra of the two stars encircled by boxes
in Fig.~\ref{fig:integrated-flux}, clearly the spectra recovered with
$\Fjoint(\V{x})$ (thin curves marked with open triangles) are of much
higher quality than the spectra estimated when treating the spectral channels
independently (thin curves marked with open boxes).  Compared to the true
spectra (thick lines) there is however a small but significant bias in the
spectra obtained with $\Fjoint(\V{x})$.  This is not unexpected as the
mixed norm implemented by $\Fjoint(\V{x})$ results in an attenuation as
shown by \Eq{eq:joint-sparse-prox}.

As shown by Fig.~\ref{fig:detect}, in the reconstructed gray image or in the
image reconstructed with $\Fjoint(\V{x})$, all sources whose mean flux is
greater than 1 are true positive detections while all false positives have a
smaller mean flux.  We therefore select the sources with mean fluxes greater
than this level to apply the debiasing method described in
Sec.~\ref{sec:debiasing} to effectively remove this bias as shown by the thin
curves with filled triangles in Fig.~\ref{fig:spectra}.

\begin{figure}
  \centering
  \includegraphics[width=84mm]{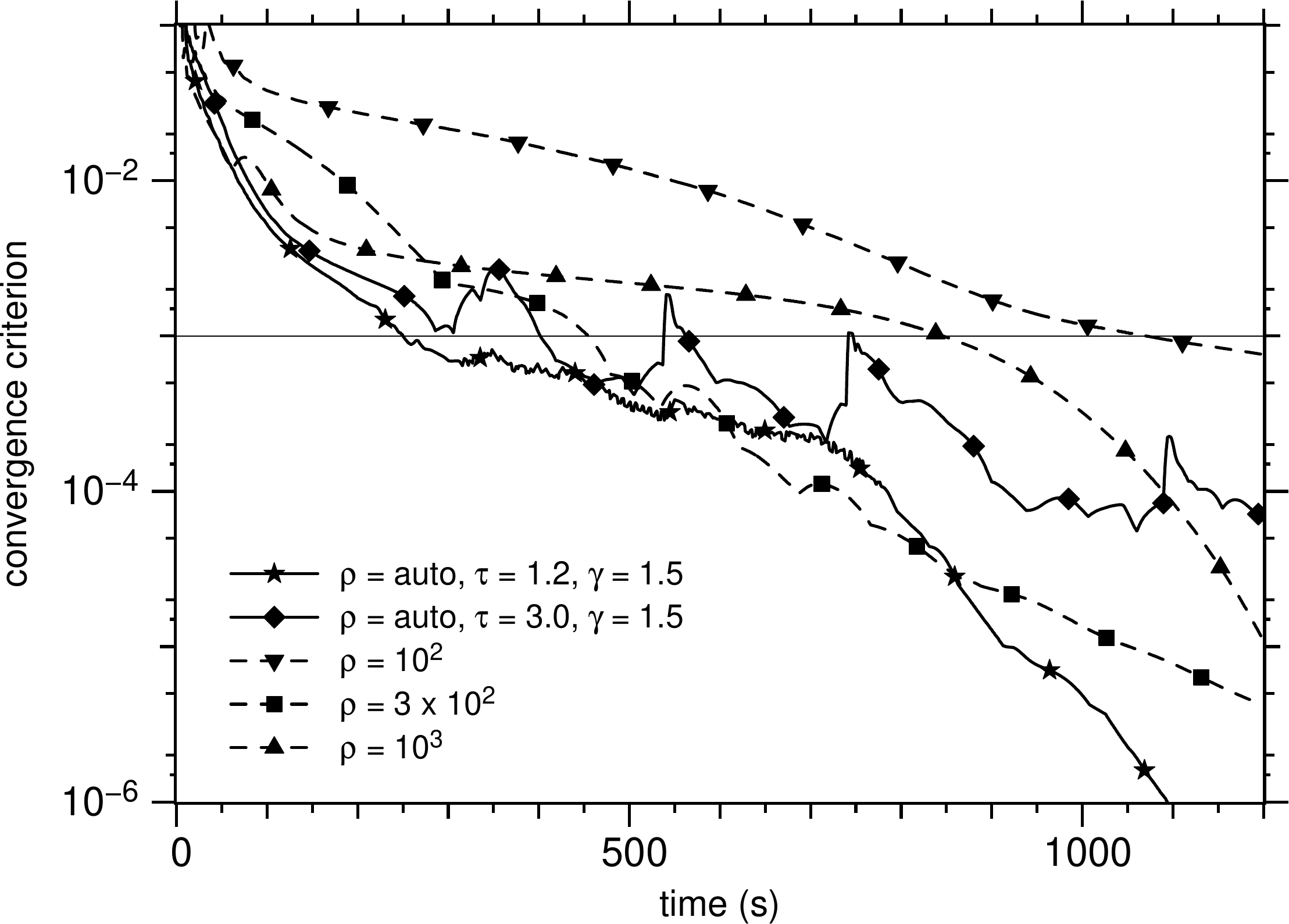}
  \caption{Evolution of the convergence criterion $\phi^{(t)}$, defined in
  \Eq{eq:convergence-criterion-alt}, for different strategies to
    choose the augmented penalty parameter. Dashed curves are for a constant
    $\rho$.  Solid curves are for $\rho$ automatically set to have
    $\eta^{(t)}_\Tag{alt} \approx 1$. \label{fig:convergence-criterion}}
\end{figure}

In our reconstructions with the regularization $\Fjoint(\V{x})$, we found that
$\phi^{(t)} < 10^{-3}$ was a good threshold for the global convergence of the
algorithm and we compared the different strategies proposed in
Sec.~\ref{sec:tuning-rho} to set the augmented penalty parameter $\rho$.  The
evolution of the convergence criterion for some of these strategies is plotted
in Figure~\ref{fig:convergence-criterion}.  With a constant value for $\rho$,
we observed that the rate of convergence is quite sensitive to the value of
the augmented penalty parameter.  Indeed with $\rho=3\times10^{2}$ (which is
the best value we found), the algorithm converged in 455\,s, while it took
1\,071\,s and 847\,s with $\rho=10^{2}$ and $\rho=10^{3}$ respectively.
Although we did not try many different values for the parameters $\tau$ and
$\gamma$, we found that the automatic strategies for setting $\rho$ with
Algorithm~2 and $\eta^{(t)}$ or according to \Eq{eq:rho-forward-update} failed
with their convergence criterion oscillating with $\phi^{(t)} \approx
10^{-2}$.  In fact, in spite of the loss of time due to the number of retries
needed to find a correct value for $\rho$ at each ADMM iteration, we found
that the best strategy was to use $\eta^{(t)}_\Tag{alt}$ in Algorithm~2.  In
this case, it seemed to be better to use a tighter tolerance for
$\eta^{(t)}_\Tag{alt} \approx 1$ as with $\tau=3$ the algorithm converged in
402\,s, while it took only 250\,s with $\tau=1.2$ (see
Fig.~\ref{fig:convergence-criterion}).

\section{Discussion and Perspectives}

We have shown the importance of using trans-spectral constraints to improve
the quality of the restoration of the multi-spectral brightness distribution
$I_\lambda(\V{\theta})$ of an astronomical target from optical interferometric
data.  These results confirm what has been observed for other types of data
(like integral field spectroscopy).

For the moment, our demonstration is restricted to specific objects which are
spatially sparse (\eg point-like sources) and must be generalized to other
types of spatial distributions.  Being implemented by non-differentiable cost
functions, spatial sparsity requires specific optimization algorithms.  We
have shown that variable splitting by the alternate direction method of
multipliers (ADMM) is suitable to solve the optimization problem in a short
amount of time.  In addition to being able to deal with non-differentiable
criteria, the ADMM method leads to splitting the full problem in sub-problems
that are easier to solve and that may be independent.  This straightforwardly
gives the opportunity of speeding up the code, \eg by means of
parallelization.  This possibility remains if other priors are used, \eg to
account for a smooth spatial distribution.

To simplify the problem at hand, we considered that complex visibilities have
been measured.  At optical wavelengths, this is only possible with phase
referencing \cite{Delplancke_at_al-2003-Prima}.  In order to process most
existing interferometric data, we will have to modify the likelihood term
$\Fdata(\V{x})$ and use a non-linear method (\ie not the linear conjugate
gradients) to update the auxiliary variables $\V{z}$.  The new algorithm that
we proposed, because it splits the two cost functions, $\Fdata(\V{z})$ and
$\Fprior(\V{x})$, may however be an efficient alternative to the variable
metric method used in \Mira \cite{Thiebaut-2008-Marseille,
  Thiebaut-2002-optim_bdec} or the non-linear conjugate gradients method in
\BSMEM \cite{Buscher-1994-BSMEM, Baron_Young-2008-Marseille} which considers
directly the sum of the cost functions.


\section*{Acknowledgments}

This work is supported by the French ANR (\emph{Agence Nationale de la
  Recherche}), Éric Thiébaut and Lo\"ic Denis work for the MiTiV project
(\emph{Méthodes Inverses de Traitement en Imagerie du Vivant},
ANR-09-EMER-008) and Ferréol Soulez is funded by the POLCA project
(\emph{Percées astrophysiques grâce au traitement de données
  interférométriques polychromatiques}, ANR-10-BLAN-0511).

Our algorithm has been implemented and tested with \Yorick
(\myURL{http://yorick.sourceforge.net/}) which is freely available.


\appendix

\section{Proximity Operators for Spatial Sparsity of Non-Negative Variables}
\label{sec:proximity-operators}

Updating of the variables $\V{x}$ by \Eq{eq:x-update} and
(\ref{eq:x-proxy-def}) in the ADMM method consists in solving a problem of the
form:
\begin{equation}
  \label{eq:admm-variable-update}
  \min_{\V{x} \in \Set{X}}
  \left\{\alpha\,f(\V{x})
  + \frac{1}{2} \, \Norm{\V{x} - \proxy{\V{x}}}_2^2\right\}
  \, ,
\end{equation}
with $\alpha = \mu/\rho^{(t)} > 0$ and $f(\V{x}) = \Fprior(\V{x})$.  Solving
problem (\ref{eq:admm-variable-update}) is very close to applying the
so-called proximity operator (also known as \emph{Moreau proximal mapping}) of
the function $\alpha\,f(\V{x})$ which is defined by
\cite{Combettes_Pesquet-2011-proximal_methods}:
\begin{equation}
  \label{eq:prox-def}
  \prox_{\alpha\,f}(\proxy{\V{x}})
  \bydef \argmin_{\V{x} \in \Reals^N}
  \left\{\alpha\,f(\V{x}) + \frac{1}{2}\,\Norm{\V{x} - \proxy{\V{x}}}_2^2\right\}
  \, .
\end{equation}
Proximity operators for non differentiable cost functions like
$\Fsparse(\V{x})$ or $\Fjoint(\V{x})$ have already been derived
\cite{Combettes_Pesquet-2011-proximal_methods} and we simply need to modify
them to account for the additional constraint that $\V{x} \in \Set{X}$.  Since
$\Set{X}$ is the subset of non-negative vectors of $\Reals^N$, \ie $\Set{X} =
\Reals^{N}_{+}$, we denote by:
\begin{equation}
  \prox_{\alpha\,f}^{+}(\proxy{\V{x}}) \bydef
  \argmin_{\V{x} \in \Reals^{N}_{+}}
  \left\{\alpha\,f(\V{x}) + \frac{1}{2}\,\Norm{\V{x} - \proxy{\V{x}}}_2^2\right\}
  \, ,
  \label{eq:prox-plus-def}
\end{equation}
the modified proximity operator that we use to update the variables $\V{x}$ in
our algorithm while accounting for non-negativity.

\subsection{Separable Sparsity}

The proximity operator for $\Fsparse(\V{x})$ defined in \Eq{eq:sparse-prior},
that is the $\ell_1$-norm of $\V{x}$, is the so called \emph{soft
  thresholding} operator \cite{Combettes_Pesquet-2011-proximal_methods}:
\begin{equation}
  \prox_{\alpha\,\Fsparse}\!(\proxy{\V{x}})_{n,\ell}
  = \left\{
  \begin{array}{ll}
    \proxy{x}_{n,\ell} - \alpha
    & \text{if\ } \proxy{x}_{n,\ell} > \alpha\,; \\[1ex]
    \proxy{x}_{n,\ell} + \alpha
    & \text{if\ } \proxy{x}_{n,\ell} < -\alpha\,; \\[1ex]
    0 & \text{else.}
  \end{array}
  \right.
\end{equation}
Imposing the  non-negativity is straightforward and yields:
\begin{equation}
  \prox_{\alpha\,\Fsparse}^{+}\!(\proxy{\V{x}})_{n,\ell}
  = \left\{
  \begin{array}{ll}
    \proxy{x}_{n,\ell} - \alpha
    & \text{if\ } \proxy{x}_{n,\ell} > \alpha\,; \\[1ex]
    0 & \text{else.}
  \end{array}
  \right.
  \label{eq:sparse-prox-plus}
\end{equation}
This shows that if $\alpha \ge \max_{n,\ell} \proxy{x}_{n,\ell}$, the output
of the proximity operator is zero everywhere.

\subsection{Spatio-Spectral Regularization}

Using the joint-sparsity regularization $\Fjoint(\V{x})$ given by
\Eq{eq:joint-sparse-prior}, we aim at minimizing the criterion:
\begin{displaymath}
  c(\V{x}) =
  \alpha\,\underbrace{
    \sum\nolimits_{n}
  \left(\sum\nolimits_{\ell} x_{n,\ell}^2\right)^{\!1/2}
  }_{\displaystyle \Fjoint(\V{x})}
  + \frac{1}{2}\,\Norm{\V{x} - \proxy{\V{x}}}_2^2
\end{displaymath}
which is strictly convex with respect to the variables $\V{x}$
\cite{Fornasier_Rauhut-2008-joint_sparsity}.  We note that $c(\V{x})$ is
separable with respect to the pixel index $n$.  Thus all computations can be
done independently for the spectral energy distribution of each pixel.

Considering first the unconstrained case and for variables $\V{x}$ such that
the function $\Fjoint(\V{x})$ is differentiable, minimizing $c(\V{x})$ with
respect to the variables $\V{x}$ amounts to finding the root of the partial
derivatives of $c(\V{x})$:
\begin{align}
  \PDer{c(\V{x})}{x_{n,\ell}} = 0
  & \quad\Longleftrightarrow\quad
  \frac{\alpha}{\beta_n}\,x_{n,\ell} +
  \left(x_{n,\ell} - \proxy{x}_{n,\ell}\right) = 0 \notag\\
  & \quad\Longleftrightarrow\quad
  x_{n,\ell} = \frac{\proxy{x}_{n,\ell}}{1 + \alpha/\beta_n} \, ,
  \label{eq:joint-sparse-intermediate-solution}
\end{align}
with:
\begin{equation}
  \beta_n \bydef \left(\sum\nolimits_{\ell} x_{n,\ell}^2\right)^{\!1/2} \, ,
  \label{eq:beta-def}
\end{equation}
the Euclidean norm of the spectral energy distribution of the $n$th pixel of
$\V{x}$.  Assuming for the moment that $\beta_n > 0$, otherwise $c(\V{x})$ is
not differentiable, combining \Eq{eq:joint-sparse-intermediate-solution} and
\Eq{eq:beta-def} yields:
\begin{equation}
  \beta_n = \frac{\proxy{\beta}_n}{1 + \alpha/\beta_n} \, ,
  \label{eq:beta-intermediate-solution}
\end{equation}
since $\alpha/\beta_n > 0$ and with:
\begin{equation}
  \proxy{\beta}_n \bydef
  \left(\sum\nolimits_{\ell} \proxy{x}_{n,\ell}^2\right)^{\!1/2} \, .
  \label{eq:proxy-beta-def}
\end{equation}
Solving \Eq{eq:beta-intermediate-solution} for $\beta_n$
yields:
\begin{equation}
  \beta_n = \proxy{\beta}_n - \alpha \, .
  \label{eq:beta-solution}
\end{equation}
The non-differentiable case occurs when the above expression yields a value of
$\beta_n$ which is not strictly positive, that is when $\proxy{\beta}_n \le
\alpha$, in which case the minimum of the cost function is given by
$x_{n,\ell} = 0, \forall \ell$.  Finally, the proximity operator of
$\alpha\,\Fjoint(\V{x})$ is:
\begin{equation}
  \prox_{\alpha\,\Fjoint}(\proxy{\V{x}})_{n,\ell}
  = \left\{
  \begin{array}{ll}
    \displaystyle
    \left(1 - \frac{\alpha}{\proxy{\beta}_n}\right)\,\proxy{x}_{n,\ell}
    & \text{if\ } \proxy{\beta}_n > \alpha \, ; \\[1ex]
    0 & \text{else}
  \end{array}
  \right.
  \label{eq:joint-sparse-prox}
\end{equation}
where $\proxy{\beta}_n$ is the Euclidean norm of the spectral energy
distribution of the $n$th pixel of $\proxy{\V{x}}$ defined by
\Eq{eq:proxy-beta-def}.

In the differentiable case, requiring that $x_{n,\ell} \ge 0$ yields a simple
modification of the unconstrained solution given by
\Eq{eq:joint-sparse-intermediate-solution}:
\begin{equation}
  x_{n,\ell} = \frac{\max(0, \proxy{x}_{n,\ell})}{1 + \alpha/\beta_n} \, ,
\end{equation}
since $1 + \alpha/\beta_n > 0$. The rest of the reasoning is similar than the
unconstrained case except that $\proxy{\beta}_n$ has to be replaced by
$\proxy{\beta}_n^+$ the Euclidean norm of the spectral energy distribution of
the $n$th pixel of $\max(0,\proxy{\V{x}})$.  The proximity operator of
$\Fjoint(\V{x})$ modified to account for non-negativity is finally:
\begin{equation}
  \prox_{\alpha\,\Fjoint}^{+}(\proxy{\V{x}})
  = \prox_{\alpha\,\Fjoint}\bigl(\max(0,\proxy{\V{x}})\bigr) \, .
  \label{eq:joint-sparse-prox-plus}
\end{equation}

\section{Stopping criterion for the conjugate gradient method}
\label{sec:Eckstein-Bertsekas-constraint}

We derive here a possible strategy to set the stopping criterion for the
conjugate gradient method used to update the auxiliary variables $\V{z}$ so as
to guarantee the global convergence of the ADMM method.  If
$\V{z}^{(t)}_\Tag{exact}$ is the solution of the linear system
$\M{A}^{(t)}\cdot\V{z} = \V{b}^{(t)}$, then:
\begin{displaymath}
  \M{A}^{(t)}\cdot\V{z}^{(t)}_\Tag{exact} - \V{b}^{(t)} = \V{0}\,;
\end{displaymath}
while, for the approximate solution:
\begin{displaymath}
  \M{A}^{(t)}\cdot\V{z}^{(t)} - \V{b}^{(t)} = \V{v}^{(t)}\,,
\end{displaymath}
where $\V{v}^{(t)}$ are the so called residuals at the end of the conjugate
gradients iterations.  Then:
\begin{displaymath}
  \V{z}^{(t)} - \V{z}^{(t)}_\Tag{exact} =
  \bigl[\M{A}^{(t)}\bigr]^{-1} \cdot \V{v}^{(t)} \, .
\end{displaymath}
As $\M{A}^{(t)} = \M{H}\T\!\!\cdot\M{W}\cdot\M{H} + \rho^{(t)} \, \M{I}$ with
$\rho^{(t)} > 0$ and since $\M{H}\T\!\!\cdot\M{W}\cdot\M{H}$ is at least
positive semi-definite, the smallest eigenvalue of $\M{A}^{(t)}$ is greater or
equal $\rho^{(t)}$, thus:
\begin{displaymath}
  \bigNorm{\V{z}^{(t)} - \V{z}^{(t)}_\Tag{exact}}_2 \le
  \bigNorm{\V{v}^{(t)}}_2/\rho^{(t)} \, .
\end{displaymath}
To have:
\begin{displaymath}
  \sum_{t = 1}^{\infty} \bigNorm{\V{z}^{(t)} - \V{z}^{(t)}_\Tag{exact}}_2 < \infty
\end{displaymath}
a sufficient condition is therefore to make sure that:
\begin{displaymath}
  \sum_{t = 1}^{\infty} \bigNorm{\V{v}^{(t)}}_2/\rho^{(t)} < \infty \, .
\end{displaymath}
This can be achieved by imposing at each iteration that the stopping criterion
for the conjugate gradients be such that:
\begin{equation}
  \label{eq:cg-conv}
  \bigNorm{\V{v}^{(t)}}_2 \le \gamma_\Tag{CG}\,\rho^{(t)}\,\xi_\Tag{CG}^t
\end{equation}
with $\gamma_\Tag{CG} > 0$ and $\xi_\Tag{CG} \in (0,1)$ since then:
\begin{align*}
  \sum_{t = 1}^{\infty} \bigNorm{\V{z}^{(t)} - \V{z}^{(t)}_\Tag{exact}}_2
  &\le \sum_{t = 1}^{\infty} \bigNorm{\V{v}^{(t)}}_2/\rho^{(t)} \\
  &\le \gamma_\Tag{CG} \, \sum_{t = 1}^{\infty} \xi_\Tag{CG}^t
   = \frac{\gamma_\Tag{CG} \, \xi_\Tag{CG}}{1 - \xi_\Tag{CG}} \, .
\end{align*}
Note that the sum is finite if and only if $\abs{\xi_\Tag{CG}} < 1$.



\end{document}